\documentclass[journal,comsoc]{IEEEtran}
\usepackage[T1]{fontenc}% optional T1 font encoding

\ifCLASSINFOpdf
\else
\fi 
\usepackage{graphicx}
\usepackage[numbers,sort&compress]{natbib}
\usepackage{amsmath,amssymb,amsfonts}
\usepackage{algorithm,algorithmic}
\usepackage{subfigure}
\usepackage{textcomp}

\DeclareMathOperator*{\argmin}{arg\,min}

\usepackage{pifont}

\usepackage{amsmath}
\usepackage[cmintegrals]{newtxmath}
\begin{document}
\title{DynaComm: Accelerating Distributed CNN Training between Edges and Clouds through Dynamic Communication Scheduling}

\author{Shangming~Cai,
        Dongsheng~Wang,~\IEEEmembership{Member,~IEEE,}
        Haixia~Wang,~\IEEEmembership{Member,~IEEE,}
        Yongqiang~Lyu,~\IEEEmembership{Member,~IEEE,}
        Guangquan~Xu,~\IEEEmembership{Member,~IEEE,}
        Xi~Zheng,~\IEEEmembership{Member,~IEEE,}
        and~Athanasios~V.~Vasilakos,~\IEEEmembership{Senior~Member,~IEEE}% <-this % stops a space
\thanks{Manuscript received March 1, 2021; revised July 25, 2021; accepted August 28, 2021. Date of publication September 15, 2021; date of current version September 15, 2021. (Corresponding author: Yongqiang Lyu and Guangquan Xu.)}
\thanks{Shangming Cai is with the Department of Computer Science and Technology, Tsinghua University, Beijing 100084, China (e-mail:csm16@mails.tsinghua.edu.cn).}%
\thanks{Dongsheng Wang is with the Department of Computer Science and Technology, Tsinghua University, Beijing 100084, China, with the Beijing National Research Center for Information Science and Technology, Tsinghua University, Beijing 100084, China, and with the Cyberspace Security Research Center, Peng Cheng Laboratory, Shenzhen 518066, China (e-mail:wds@tsinghua.edu.cn).}%
\thanks{Haixia Wang and Yongqiang Lyu are with the Beijing National Research Center for Information Science and Technology, Tsinghua University, Beijing 100084, China (e-mail:hx-wang@tsinghua.edu.cn;luyq@tsinghua.edu.cn).}%
\thanks{Guangquan Xu is with the Big Data School, Qingdao Huanghai University, Qingdao 266427, China, and with the Tianjin Key Laboratory of Advanced Networking (TANK), College of Intelligence and Computing, Tianjin University, Tianjin 300350, China (e-mail:losin@tju.edu.cn).}%
\thanks{Xi Zheng is with the Department of Computing, Macquarie University, Sydney, NSW 2109, Australia (e-mail:james.zheng@mq.edu.au).}%
\thanks{Athanasios V. Vasilakos is with the College of Mathematics and Computer Science, Fuzhou University, Fuzhou 350116, China, with the School of Electrical and Data Engineering, University of Technology Sydney, Australia, and with the Department of Computer Science, Electrical and Space Engineering, Lulea University of Technology, Lulea 97187, Sweden (e-mail:th.vasilakos@gmail.com).}}%

\maketitle

\begin{abstract}
To reduce uploading bandwidth and address privacy concerns, deep learning at the network edge has been an emerging topic. Typically, edge devices collaboratively train a shared model using real-time generated data through the Parameter Server framework. Although all the edge devices can share the computing workloads, the distributed training processes over edge networks are still time-consuming due to the parameters and gradients transmission procedures between parameter servers and edge devices. Focusing on accelerating distributed Convolutional Neural Networks (CNNs) training at the network edge, we present DynaComm, a novel scheduler that dynamically decomposes each transmission procedure into several segments to achieve optimal layer-wise communications and computations overlapping during run-time. Through experiments, we verify that DynaComm manages to achieve optimal layer-wise scheduling for all cases compared to competing strategies while the model accuracy remains untouched.
\end{abstract}

\begin{IEEEkeywords}
Edge computing, deep learning training, dynamic scheduling, convolutional neural network.
\end{IEEEkeywords}

% \fi
\IEEEpeerreviewmaketitle

\section{Introduction and Motivation}

\IEEEPARstart{D}{eep} learning models like Convolutional Neural Networks (CNNs) have been broadly used for a wide range of applications at the edge such as Face Recognition, Object Detection, and Video Surveillance \cite{DBLP:conf/mobicom/ZhangCBJB15,DBLP:conf/edge/HungABGYBP18}. Enabled by such applications, concepts like autonomous driving \cite{DBLP:conf/percom/DengZZCLK20,DBLP:conf/asplos/LinZHSHTM18}, smart home, and smart city \cite{DBLP:conf/iccv/OuyangW13}, which are driven by the vision of Internet of Things (IoT), are becoming a reality in recent years \cite{DBLP:journals/iotj/XuWJLLZLXG20,DBLP:journals/iotj/BianchiBLFMM19,DBLP:journals/tetci/Chen0WDWZ019}. However, as a data-driven technology, deep learning commonly requires a lot of computation power and huge datasets. Not only this, user data privacy issues and network bandwidth consumption are critical as well in such scenarios \cite{DBLP:journals/wc/NiZV21}. To address these concerns, deploying deep learning training to the network edge \cite{zhang2020achieving,DBLP:journals/comsur/MaoYZHL17,DBLP:journals/iotj/ShiCZLX16}, which offloads computation tasks to the edge devices and avoids dataset uploading, has become a popular topic.

Typically, globally shared parameters are stored on the cloud servers (denoted as parameter servers) while the datasets are produced on many edge devices (i.e., workers), which is shown in Fig. 1. Each edge device pulls up-to-date parameters from the servers to perform local training and remote updating either synchronously or asynchronously, which can be described as the Parameter Server (PS) framework \cite{DBLP:conf/osdi/LiAPSAJLSS14,DBLP:journals/corr/ChenLLLWWXXZZ15}. Guaranteed by this, the data owners do not have to upload their sensitive information to the cloud, which preserves their privacy and saves a lot of network bandwidth by keeping data local to the trusted network
edge.

Although the time-consuming dataset uploading processes are avoided, the introduced communication procedures (i.e., parameter transmissions and gradient transmissions) between parameter servers and edge devices have become a new challenge to be addressed. Normally, these communication procedures take a certain amount of time to complete and depend on the network condition. If these tensor transmission procedures are delayed due to the slow network, the training process will be significantly bottlenecked. This phenomenon may be even more prominent when the latency is high or the edge network bandwidth is saturated.

To overcome this drawback, several studies have illuminated a path that the communication and computation procedures can be further decomposed into mini-procedures layer-by-layer to hide the communication overheads in computations. For instance, Poseidon \cite{DBLP:conf/usenix/ZhangZXDHLHWXX17} adopts a layer-by-layer transmission strategy to enable wait-free backward propagation. Briefly, each layer's gradient transmission mini-procedure is launched once the backward computation of this layer is completed. In the meanwhile, the computation devices will continue to process the remaining backward computation in parallel. By hiding the communication overheads and reducing the idle time on edge devices, the training process can be accelerated.

However, the layer-by-layer transmission strategy fails to achieve optimal scheduling for it neglects that each independent mini-procedure introduces extra function calls and inter-node coordination overheads, especially over a slower edge network. To address this issue, iBatch \cite{DBLP:conf/aaai/WangP019} adopts a greedy tactic to selectively batch the parameter transmission mini-procedures to avoid unnecessary decomposition overheads and achieve better overlapping in the forward propagation. As for backward propagation, iPart \cite{DBLP:journals/tpds/WangPZWX21} (i.e., the extended version of iBatch) covers this part recently based on the same intuition. However, the performance improvement brought by iBatch (or iPart) is limited since the greedy tactic applied leads to a locally optimal solution in most cases. Due to this, iBatch even shows poor performance in some cases compared to the vanilla layer-by-layer transmission strategy.

Motivated by these concerns, our goals are (1) to achieve optimal layer-wise communications and computations overlapping during run-time (2) to minimize the total execution time of each iteration for CNN training over edge networks. To achieve these two goals, we need to design a general-purpose communication scheduler for both forward and backward propagations. In this paper, we formulate the communication scheduling as a Zero-One Integer Programming problem \cite{pierce1968application}. Based on this, we then present a novel layer-wise communication scheduler, DynaComm, which is mainly powered by a lightweight real-time profiling module and two neat Dynamic-Programming (DP) algorithms for forward and backward propagation scheduling with $\mathcal{O}(L^3)$ time complexity. Through a series of experiments, we verify that DynaComm manages to achieve optimal layer-wise communication scheduling compared to the aforementioned competing methods.

Our contributions are summarized below:

\begin{itemize}
\item We discuss a communication scheduling problem, which is the underlying reason for the longer iteration execution time when performing deep learning at the network edge.
\item We present a novel general-purpose communication scheduler, DynaComm, to address the layer-wise scheduling problem for both the parameters and the gradients communications for CNN training over edge networks.
\item We conduct extensive experiments to validate the efficiency of DynaComm. The results show that DynaComm manages to achieve optimal layer-wise scheduling compared to competing strategies, and it reduces the running time of each iteration by up to 41.92\% while the model accuracy remains untouched.
\item We also discuss the sensitivity of DynaComm to the computation/communication ratio and verify the time complexity of DynaComm experimentally.
\end{itemize}

The rest of this paper is organized as follows. Section II shows the related work. Section III formulates the problem. Section IV introduces the overall design of DynaComm in detail. Section V demonstrates the implementation and experiment methodology along with the results and analyses. Section VI then discusses the applicability and limitation of DynaComm. And Section VII summarizes the paper in the end.

\begin{figure}[t]
\centering
\includegraphics[scale=0.3]{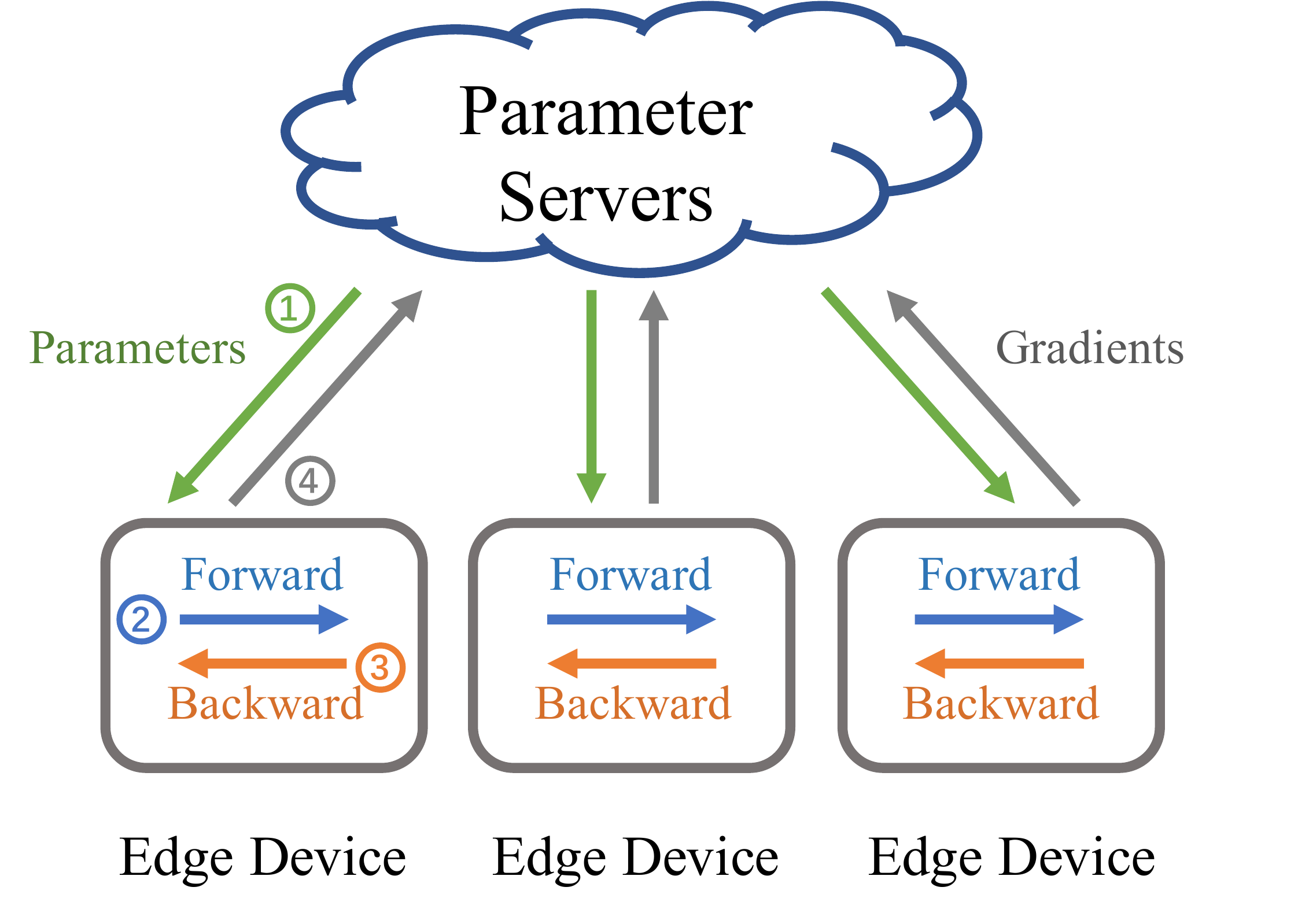}
\caption{The abstraction of the distributed learning framework at the network edge. Typically, every iteration at each edge device includes four procedures arranged in sequence: \ding{172} Parameter transmission, \ding{173} Forward propagation, \ding{174} Backward propagation, and \ding{175} Gradient transmission.}
\label{fig:arch}
\end{figure}

\section{Related Work}
In this section, we briefly review and discuss some recent works related to deep learning at the network edge and the communication optimization for distributed learning.

\subsection{Deep Learning with Edge Computing} Edge computing, which pushes computation tasks to the nodes that are placed close to end devices, is a viable way to meet the high computation and low-latency requirements of deep learning on edge devices. Apart from this, it also provides additional benefits in terms of privacy and bandwidth efficiency \cite{DBLP:conf/ccs/ShokriS15}. But that comes at the price of making communication costs become a major concern. Furthermore, the limited computing capability and the energy consumption \cite{DBLP:conf/icpads/GuKMK19} of the edge device could also be troublesome problems. To address such challenges, a lot of research has focused on performance optimization for this emerging topic. For example, Neurosurgeon \cite{DBLP:conf/asplos/KangHGRMMT17} proposes a system that can automatically partition DNN between the mobile device and cloud at layer level to achieve lower latency and less mobile energy consumption. Due to this technique, the intermediate results can be sent to an edge server once the first few layers of the DNN model have been computed. Li et al. \cite{DBLP:journals/network/LiOD18} and Eshratifar et al. \cite{DBLP:journals/tmc/EshratifarAP21} also propose similar methods that divide the DNN model into the cloud segment and the edge segment. The edge server computes the initial layers of the DNN model while the cloud computes the higher layers of the DNN. Such design utilizes the computation capabilities of both the edge server and the cloud, and it reduces the network traffic between the edge server and the cloud. Moreover, MoDNN \cite{DBLP:conf/date/MaoCNKC17} and DeepThings \cite{DBLP:journals/tcad/ZhaoBG18} both adopt fine-grained partitioning to distribute DNN executions on lightweight end devices based on the computation capabilities and the memory usage of the end devices during run-time. To address the collaborative training in a hierarchical manner, HierTrain \cite{DBLP:journals/ojcs/LiuCZL20} presents a hybrid parallelism method to adaptively assign the model layers and the data samples across edge devices, edge servers, and the cloud.

\subsection{Communication Optimization for Distributed Learning}

To improve system utilization and accelerate the training procedure in distributed DNN training, several tensor-wise scheduling techniques have been proposed in recent years. As an example, TicTac \cite{hashemi2018tictac} provides two priority-based heuristics, TIC and TAC, to achieve near-optimal scheduling of computation and communication at the operation level. Similarly, Geryon \cite{DBLP:conf/infocom/WangLG20} leverages multiple flows with different priorities to transfer parameters of different urgency levels. It also offloads the scheduling logic to the NIC hardware and prioritizes the urgent parameter transfers in the entire network fabric. Based on parameter slicing techniques, P3 \cite{jayarajan2019priority}, a priority-based synchronization mechanism, is presented to conduct the scheduling at a finer granularity. But in the meanwhile, it introduces a tricky parameter (i.e., the slicing granularity) to be tuned. To address this issue, ByteScheduler \cite{DBLP:conf/sosp/PengZCBYLWG19} introduces a Bayesian Optimization approach to auto-tune the slicing granularity during its tensor-level scheduling. All studies above do not pay much attention to the introduced overheads, which may lead to deviations and diminish the performance gains to some extent, especially for cross-regional distributed training such as distributed edge learning.

Some compression-based techniques like Gradient Quantization \cite{DBLP:conf/nips/WenXYWWCL17,DBLP:conf/nips/AlistarhG0TV17} and Gradient Sparsification \cite{DBLP:conf/iclr/LinHM0D18,DBLP:conf/aaai/ChenCBAZG18,DBLP:conf/nips/WangniWLZ18} can reduce the gradient communication time by cutting down the data size. However, due to lossy compression, these methods have to address a trade-off between the model accuracy and the training speed. To address bandwidth constraints for distributed training across geo-distributed data centers, Gaia \cite{DBLP:conf/nsdi/HsiehHVKGGM17} proposes a policy where updates are synchronized and pushed across different data centers only when the aggregated updates are higher than a given threshold. Similarly, this method also trades the accuracy for lower network consumption. Moreover, these techniques are targeted only to gradient transmission procedures. Another solution to address this issue for both parameter and gradient transmission procedures is to utilize advanced network hardware (e.g., RDMA NIC) to derive a more efficient network environment \cite{DBLP:journals/ijpp/JiaLJLAHWC18,DBLP:conf/hpcc/RenWZWZWHJ17}. But in the meanwhile, the system implementation budget could increase dramatically as the cluster scale grows.

Compared to these techniques above, hiding the communication overheads in computations through layer-wise communication scheduling only requires few extra resources and minimum implementation modification while the model accuracy remains untouched.

\subsection{Communication and Computation Reorganizing Techniques for Deep Learning Training} Except for the scheduling methods mentioned in Section I, there is other related literature for this topic. For example, focused on addressing the communication scheduling for decentralized distributed deep learning systems based on all-reduce architecture, PACE \cite{DBLP:conf/infocom/BaoPCW20} presents a dynamic solution to identify the best schedule and granularity of tensor communication to achieve near-optimal communication and computation overlapping. The intuition is to improve bandwidth utilization and reduce the execution time of the training DAG (i.e., Directed Acyclic Graph) through communication scheduling, which is similar to iBatch and DynaComm. Moreover, some studies have shown a way to restructure the computation to accelerate local deep learning training. For example, TASO \cite{DBLP:conf/sosp/JiaPTWZA19} employs a cost-based backtracking search to find and generate operation-wise graph substitutions for computation acceleration. Ansor \cite{DBLP:conf/osdi/ZhengJSWYHWYZSG20} uses an operator fusion algorithm to convert DNNs from common layered formats to partitioned small subgraphs. Then it identifies the best programs with evolutionary search and a learned cost model to restructure computation DAG. By altering the upper-level computation DAG representation and profiling engine, the layer-wise communication scheduling methods can be used in combination with such local computation graph optimizing techniques with minimum modification.

\section{Problem Definition}
\subsection{Layer-wise Communication Scheduling}
In the default PS, every iteration at each node includes four procedures: parameter transmission (also known as pulling), forward propagation computation, backward propagation computation, and gradient transmission (i.e., pushing). For further explanations, we denote these procedures as [$pt_i$, $fc_i$, $bc_i$, $gt_i$], where integer $i$ indicates that this is the $i$-th iteration. Due to data dependency, these procedures are conducted in sequence strictly in the default PS, which is shown in Fig. 2 (a).

Without loss of generality, assume there are $L$ layers in the targeted CNN model. Then each procedure can be further decomposed into $L$ mini-procedures sequentially according to the computation graph of this targeted network, as is shown in Fig. 2 (b). For instance, $pt_i$ can be decomposed into $L$ mini-procedures, $[pt_i^1, ..., pt_i^L]$, where $L$ is the depth of this network. Note that if the computation graph contains branches, the parameters from different branches with the same depth will be considered as one layer. As for the transformation layers with no actual parameter, such as pooling, flatten, and concatenation, they are considered as a computation portion of their previous layer.

Due to data dependency, these mini-procedures are required to satisfy these partial order relations as follows:

\begin{small}
\begin{equation}
\forall \ 0 < l \leq L,\ st(pt_i^l) \prec ed(pt_i^l) \preceq st(fc_i^l) \prec ed(fc_i^l)
\label{eq1}
\end{equation}

\begin{equation}
\forall \ 0 < l \leq L,\ st(bc_i^l) \prec ed(bc_i^l) \preceq st(gt_i^l) \prec ed(gt_i^l)
\label{eq2}
\end{equation}

\begin{equation}
\forall \ 0 < l \leq L,\ ed(fc_i^l) \preceq st(bc_i^l)
\label{eq3}
\end{equation}

\begin{equation}
\forall \ 0 < m < n \leq L,\ ed(pt_i^m) \preceq st(pt_i^n)
\label{eq4}
\end{equation}

\begin{equation}
\forall \ 0 < m < n \leq L,\ ed(fc_i^m) \preceq st(fc_i^n)
\label{eq5}
\end{equation}

\begin{equation}
\forall \ 0 < m < n \leq L,\ ed(bc_i^n) \preceq st(bc_i^m)
\label{eq6}
\end{equation}

\begin{equation}
\forall \ 0 < m < n \leq L,\ ed(gt_i^n) \preceq st(gt_i^m)
\label{eq7}
\end{equation}
\end{small}where $l,m,n \in \mathbb{N}$ and function $st$ stands for the start time of this mini-procedure while function $ed$ stands for the end time. In the meanwhile, $a \prec b$ means $a$ precedes $b$ and $a \preceq b$ means $a$ precedes or equals $b$.

\begin{figure}[t]
\centering
\includegraphics[scale=0.25]{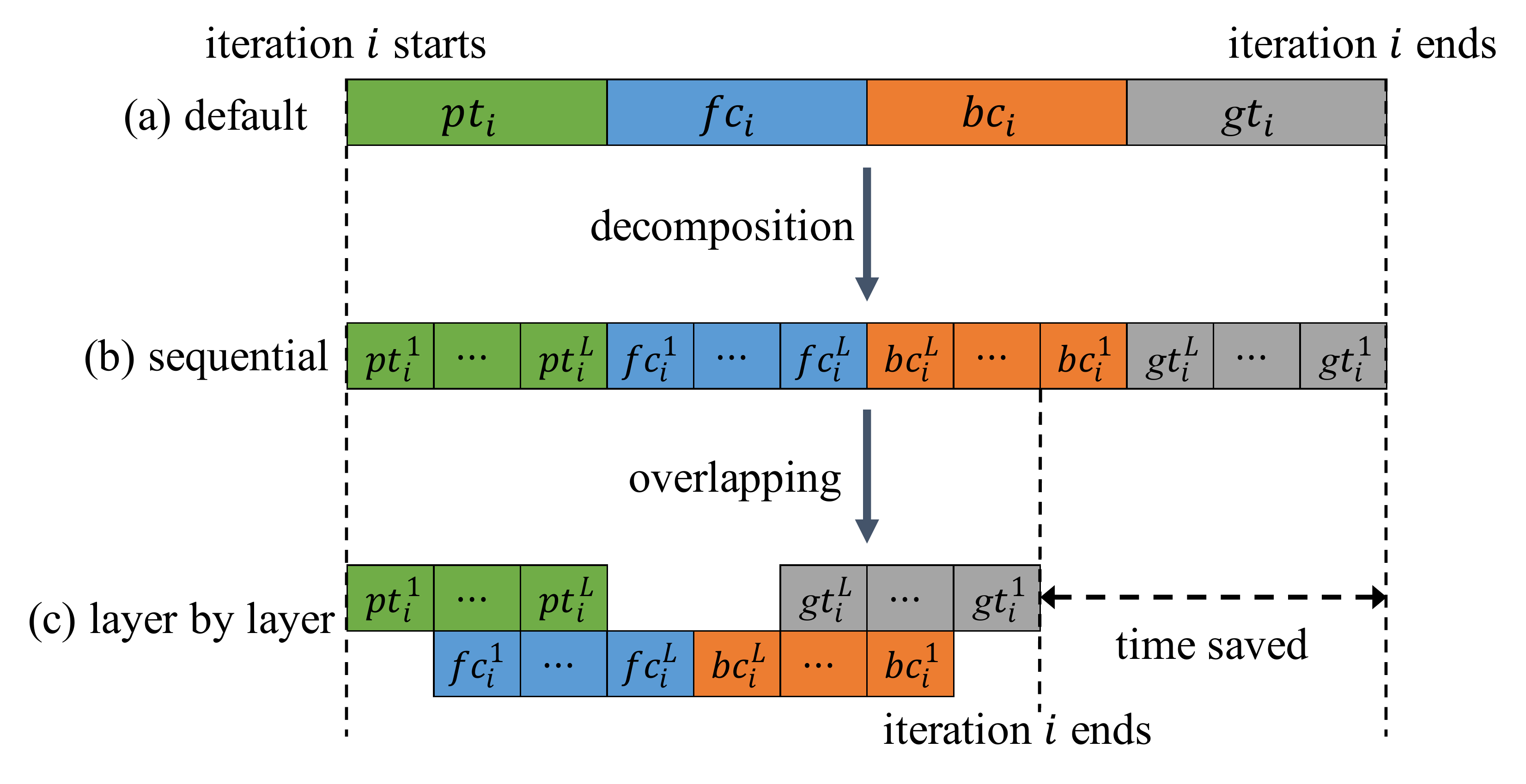}
\caption{Diagram of the procedure execution flow during iteration $i$ in the ideal case. (a) exhibits the original implementation of the default PS. (b) demonstrates that each procedure can be further decomposed into several layer-wise mini-procedures. As for (c), it shows the overlapping progress enabled by the layer-by-layer transmission strategy.}
\label{fig:ideal_overlapping}
\end{figure}

\begin{figure*}[h]
\centering
\includegraphics[scale=0.248]{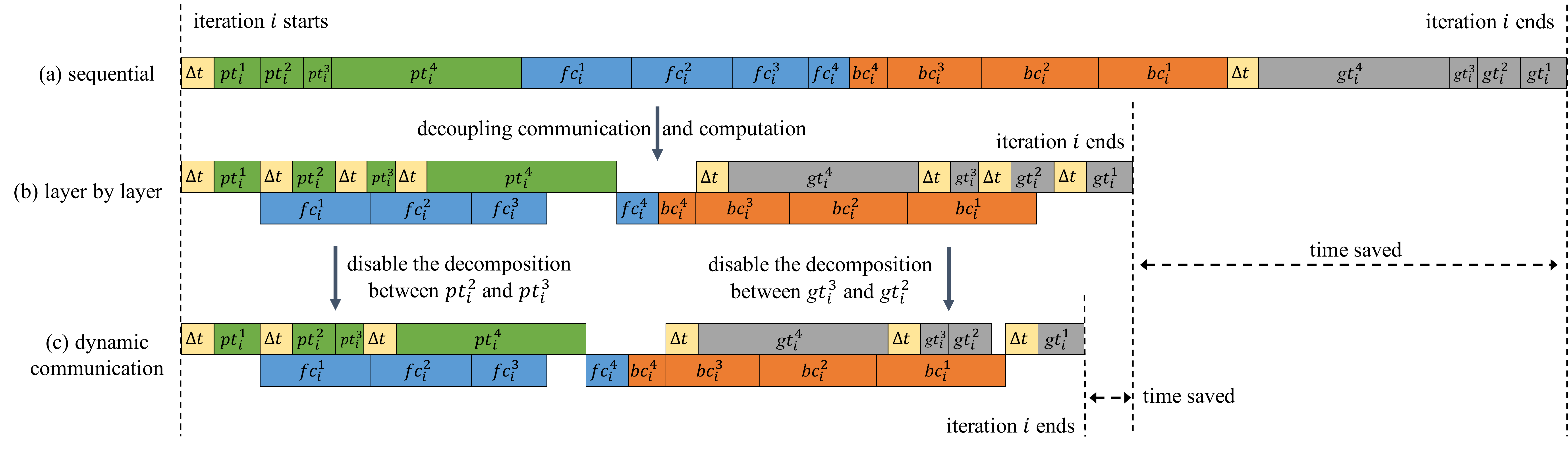}
\caption{Diagram of the procedure execution flow of a 4-layer toy network during iteration $i$ in the real circumstance. (a) exhibits the sequential implementation of the default PS. (b) demonstrates the running process of the layer-by-layer transmission strategy. As for (c), it reveals the potential performance gains that a dynamic communication scheduling strategy could bring.}
\label{fig:real_usecases}
\end{figure*}

With the help of the layer-by-layer transmission strategy, the communication and computation mini-procedures can be overlapped with each other without violating the inter-layer data dependency and the intra-layer procedure execution order. Eventually, the total running time of each iteration is reduced (exhibited in Fig. 2 (c)).

However, there remain several concerns to be addressed in real use-cases. Firstly, each independent transmission mini-procedure invokes additional overhead for the transmission set-up and extra function calls. Generally, this introduced overhead is relatively constant, which mostly depends on the system condition of the underlying infrastructure \cite{DBLP:journals/tpds/WangPZWX21}. In this study, we use $\Delta t$ to represent the expected value of the introduced overhead. Secondly, layers' properties and tensor volumes are commonly varying within a wide range in CNNs, which means layers' computation costs may not be proportional to their communication costs. For instance, the computation cost of a convolutional layer can be huge even though its communication cost is relatively small. Consequently, the layer-by-layer transmission strategy is no longer the optimal solution in the real circumstance, which poses both an opportunity and a challenge to achieve more intelligent communication scheduling, as is shown in Fig. 3.

\subsection{Problem Formulation}
In this section, we reconsider the communication scheduling as a Zero-One Integer Programming \cite{pierce1968application} problem, which relies on mutually exclusive 1 and 0 decisions to find solutions.

Firstly, we assume that there exists a decomposition position after each layer to indicate whether the parameters of the next layer should be sent with the current layer. For specification, the compulsory decomposition position after nonexistent layer $0$ is considered as the starting position while the compulsory decomposition position after the last layer $L$ is considered as the end position. Within this context, we use a sequence, $[0, 1, 2, ..., L-1, L]$, to represent the layer-by-layer transmission strategy and $[0, L]$ to represent the sequential strategy for the forward propagation. Then we define $L-1$ parameters, $p_1, p_2, ..., p_{L-1}$, where $\forall \ l \in \{1, ..., L-1\}, p_l \in\{0 ,1\}$. For other strategies in between, if the optional decomposition position $l$ is not enabled, then parameter $p_l$ equals 0. It also means that layer $(l+1)$'s parameters will be pulled along with layer $l$'s parameters. On the other hand, if this optional decomposition position $l$ is enabled, then parameter $p_l$ equals 1, and layer $(l+1)$'s parameters will be pulled by another mini-procedure. All enabled decomposition positions can be formed as a decomposition decision set $\mathcal{D}_f$.

As for the backward propagation, the compulsory decomposition position after nonexistent layer $L+1$ is considered as the starting position while the compulsory decomposition position after layer $1$ is considered as the end position. Similarly, we use $[L+1, L, ..., 2, 1]$ to represent the layer-by-layer transmission strategy and $[L+1, 1]$ to represent the sequential strategy in the default PS. Also, we define $g_1, g_2, ..., g_{L-1}$ to determine a decomposition decision $\mathcal{D}_b$ for the backward propagation, where $\forall \ l \in \{1, ..., L-1\}, g_l \in \{0, 1\}$. If the optional decomposition position after layer $(L+1-l)$ is not enabled, then parameter $g_l$ equals 0. But if this decomposition position is enabled, then parameter $g_l$ equals 1.

Here, we manage to abstract the communication scheduling problem as a Zero-One Integer Programming problem, which is given as follows:

\begin{align}
  \label{Eq:argmin}
  & \argmin\limits_{\vec{p}, \vec{g}} f_{m}( \vec{pt_i}, \vec{fc_i}, \vec{bc_i}, \vec{gt_i}, \Delta t, L, \vec{p}, \vec{g}) \\
  & \quad \text{s.t.} \quad \vec{p} = [p_1, p_2, ..., p_{L-1}]\\
  & \quad \quad \quad \ \vec{g} = [g_1, g_2, ..., g_{L-1}]\\
  & \quad \quad \quad \ \forall \ l \in \{1, ..., L-1\}, p_l \in \{0, 1\}\\
  & \quad \quad \quad \ \forall \ l \in \{1, ..., L-1\}, g_l \in \{0, 1\}
\end{align} where $\vec{pt_i}$ and $\vec{gt_i}$ are the tensor communication cost vectors while $\vec{fc_i}$ and $\vec{bc_i}$ are the computation cost vectors. These cost vectors are derived from the real-time profiler module (detailed in Section IV-A). In the meantime, $f_{m}$ is an approximate cost measurement function, which simply uses all the parameters above to estimate the total running time of an iteration with the help of the computation graph. 

Although the iteration running time can be measured by a cost measurement function $f_m$ with $\mathcal{O}(L)$ time complexity once the candidate decomposition decision and the profiling results are given, it is extremely impractical to obtain or verify an optimal solution by brute-force search. After all, the time complexity of a brute-force search is $\mathcal{O}(L \cdot 2^L)$. If the scheduling overhead is larger than the potential performance gains, then there is no point to complicate the underlying implementation.

\begin{algorithm}[tbp]
\setcounter{algorithm}{0}
 \caption{Greedy parameter scheduling of iBatch (forward)}
 \label{alg:algorithm1}
\textbf{Input}: Parameter Transmission Cost $\vec{pt_i}$, Forward Computation Cost $\vec{fc_i}$, Introduced Overhead $\Delta t$, Layer Number $L$\\
\textbf{Output}: Transmission Decomposition Decision $\mathcal{D}_f$ for Forward Propagation
 \begin{algorithmic}[1]
    \raggedright
    \STATE Derive set $S_1$ that contains all possible pairs of [$\mathcal{D}_f[1], \mathcal{D}_f[2]$]   \hfill\COMMENT{$\mathcal{D}_f[0]$ is initialized and fixed to $0$}
    \STATE From \(S_1,\) select pairs that meet \\ $\Delta t + \sum_{\mathcal{D}_f[1]+1 \leq l \leq \mathcal{D}_f[2]}pt_i^l \ge \sum_{\mathcal{D}_f[0]+1 \leq l \leq \mathcal{D}_f[1]}fc_i^l$ and form them as set $S_2$
      \STATE From $S_2$, select pairs with maximum $\sum_{\mathcal{D}_f[0]+1 \leq l \leq \mathcal{D}_f[1]}fc_i^l$
      \STATE From the selected pairs, choose the one with minimum \\ $\Delta t + \sum_{\mathcal{D}_f[0]+1 \leq l \leq \mathcal{D}_f[1]}pt_i^l$ \hfill\COMMENT{ $\mathcal{D}_f[1], \mathcal{D}_f[2]$ are set.}
  %\STATE $\mathcal{D}_f[2]$ and $\mathcal{D}_f[2]$ are settled.
  \STATE $n \leftarrow \mathcal{D}_f[1], m \leftarrow \mathcal{D}_f[2]$
  \REPEAT
   \STATE $Options$ $\leftarrow \{\}$
   \FOR{$x$ in [$m+1,L$]}
   \IF {$\Delta t + \sum_{m+1 \leq l \leq x}pt_i^l \ge \sum_{n+1 \leq l \leq m}fc_i^l$}
   \STATE $Options$.append($x$)
   \ENDIF
   \ENDFOR
    \STATE From $Options$, choose the one as $j$ that has minimum $\Delta t + \sum_{m+1 \leq l \leq j}pt_i^l - \sum_{n+1 \leq l \leq m}fc_i^l$
  \STATE $n \leftarrow m$
  \STATE $m \leftarrow j$
  \STATE ${D}_f.$append$(m)$
\UNTIL {$m = L$}
\RETURN ${D}_f$
 \end{algorithmic} 
 \end{algorithm}

\begin{algorithm}[htbp]
\setcounter{algorithm}{1}
 \caption{Greedy gradient scheduling of iBatch}
 \label{alg:algorithm2}
\textbf{Input}: Backward Computation Cost $\vec{bc_i}$, Gradient Transmission Cost $\vec{gt_i}$, Introduced Overhead $\Delta t$, Layer Number $L$\\
\textbf{Output}: Transmission Decomposition Decision $\mathcal{D}_b$ for Backward Propagation
 \begin{algorithmic}[1]
    \raggedright
    \STATE $Candidates$ $\leftarrow \{\}$
    \FOR {$n$ in [2, $L$]}
    \STATE $\mathcal{D}_{tmp} \leftarrow \{L+1\}$  \hfill\COMMENT{$\mathcal{D}_{tmp}[0]$ is initialized to $L+1$}
    \STATE $k \leftarrow 1$
    \STATE $m \leftarrow n$
    \STATE ${D}_{tmp}.$append($m$)  \hfill\COMMENT{second decomposition position}
  \REPEAT
  \STATE $Options$ $\leftarrow \{\}$
   \FOR{$x$ in [$1,m-1$]}
   \IF {$k * \Delta t + \sum_{L \geq l \geq m}gt_i^l \ge \sum_{m-1 \geq l \geq x}bc_i^l$}
   \STATE $Options$.append($x$)
   \ENDIF
   \ENDFOR
    \STATE From $Options$, choose the one as $j$ that has minimum $k * \Delta t + \sum_{L \geq l \geq m}gt_i^l - \sum_{m-1 \geq l \geq x}bc_i^l$
  \STATE ${D}_{tmp}.$append($j$)
  \STATE $m \leftarrow j$
  \STATE $k \leftarrow k + 1$
\UNTIL {$m = 1$}
\STATE $Candidates.$append($\mathcal{D}_{tmp}$)
\ENDFOR
\STATE From $Candidates$, choose the one as $\mathcal{D}_b$ that has minimum estimated total execution time

\RETURN ${D}_b$
 \end{algorithmic} 
 \end{algorithm}

\subsection{iBatch}
To address the parameter communication scheduling for the forward propagation, iBatch \cite{DBLP:conf/aaai/WangP019} uses a greedy tactic to selectively batch the transmission mini-procedures during training. Similarly, the layer-wise batching decision provided by iBatch can be represented as $\mathcal{D}_f$ since each batch is determined by two decomposition steps equivalently. In brief, two greedy algorithms are employed to make greedy batching choices at each step to maximize the overlapping of the current segment's computation and its next segment's communication. One algorithm enumerates all possible choices and does the greedy selection from the first layer to the last layer, which is detailed in Algorithm 1, while the other algorithm does the opposite, which is presented in \cite{DBLP:conf/aaai/WangP019} and omitted here. Each algorithm derives one candidate batching decision $\mathcal{D}_f$. From two candidates, iBatch chooses the one with a lower estimated total execution time to batch parameter transmission mini-procedures in the forward propagation.

As for the backward propagation, iPart \cite{DBLP:journals/tpds/WangPZWX21} (i.e., the extended version of iBatch) also presents a greedy solution based on the same intuition as iBatch. For the sake of consistency, we'll refer to iPart as a part of iBatch for the rest of this paper. The greedy gradient scheduling algorithm of iBatch is detailed in Algorithm 2. Unlike the greedy tactic for parameter transmission, iBatch uses one algorithm to generate more than two candidates through enumeration operation at line 2. Then it chooses the one with the minimum estimated total execution time to batch gradient transmission mini-procedures in the backward propagation.

However, neither the forward scheduling nor the backward scheduling problem has the greedy choice property that the locally optimal choice always leads to the globally optimum, let alone the introduced overhead $\Delta t$ changes the remaining problem as well. Due to this, iBatch even shows poor performance compared to the vanilla layer-by-layer transmission strategy in some cases.

\section{Dynamic Communication Scheduling}
In this section, we demonstrate the design detail of our novel general-purpose layer-wise communication scheduler, DynaComm. As is exhibited in Fig. 4, DynaComm is mainly built upon the original runtime dependency engine of the trainer process and powered by a lightweight real-time profiling module. To achieve optimal layer-wise communication scheduling, two neat Dynamic-Programming (DP) algorithms are proposed.

\begin{figure}[t]
\centering
\includegraphics[scale=0.42]{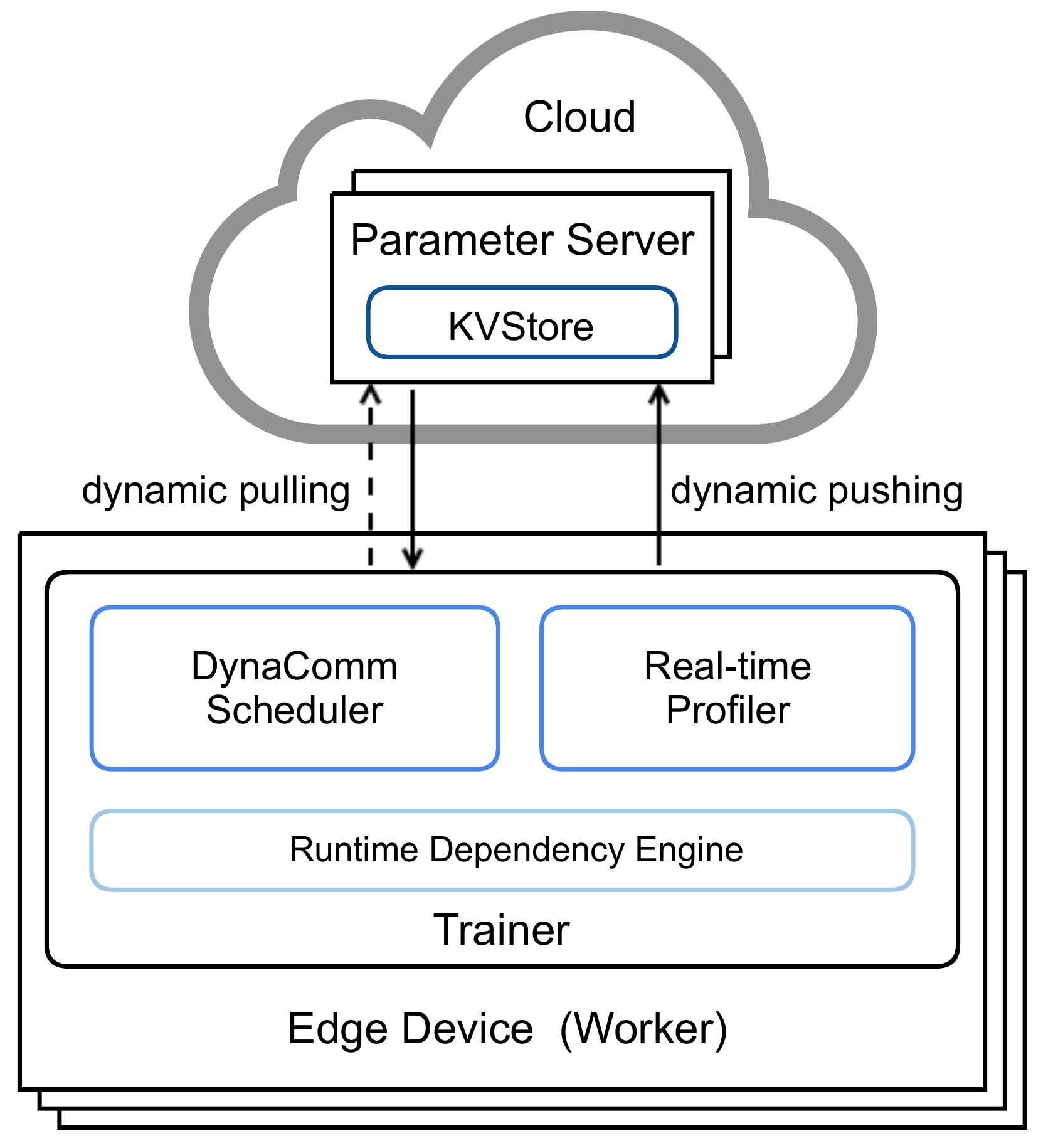}
\caption{The overview of DynaComm.}
\label{fig:arch2}
\end{figure}

\subsection{Real-Time Profiling}
To achieve dynamic communication scheduling, a lightweight profiler that reveals the real-time system status and execution condition is essential. Considering that most mainstream frameworks such as TensorFlow, PyTorch, and MXNet have already implemented a built-in profiler, DynaComm can easily exploit these profilers to do the monitoring. Through simple retrieving and reprocessing, all aforementioned cost vectors (i.e., $\vec{pt_i}, \vec{fc_i}, \vec{bc_i}, \vec{gt_i}$) and $\Delta t$ can be derived from the original profiling results (e.g., the \textit{json} file generated by mxnet.profiler). To be noticed, $\Delta t$ is assumed and considered as a constant cost value and also derived by profiling and averaging in this study.

\subsection{Dynamic Scheduling Algorithms}
Previously, we have already formulated the scheduling problem as a Zero-One Integer Programming problem. In this section, we present the principle and the design detail of DynaComm's scheduling algorithms for this problem. These two algorithms are based on the Dynamic-Programming \cite{bellman1954theory} method, which refers to simplifying a complicated problem by breaking it down into simpler sub-problems in a recursive manner.

\subsubsection{Sub-problem Definition}
As mentioned before, each enabled decomposition position (except for the end position) invokes an independent transmission mini-procedure that introduces extra overhead $\Delta t$, which changes the remaining problem. Therefore, the scheduling problem can be extremely complicated if we consider the assignment of each optional decomposition position (i.e., the decision between any two adjacent layers) as the most fine-grained sub-problem. To enable the DP algorithm, we propose and craft a subtle sub-problem definition.

Without loss of generality, we assume there are $L$ layers in our targeted network. According to definition and formulation, there exist $L-1$ optional decomposition positions and $2$ compulsory decomposition positions after layer $0$ and layer $L$ in the forward propagation. The problem is to find the optimal decomposition decision for this $L$-layer network. Here, we define its sub-problem in the forward propagation as finding the optimal decomposition decision for the first $L-1$ layers of this network, which has $L-2$ optional decomposition positions and $2$ compulsory decomposition positions after layer $0$ and layer $L-1$. Recursively, the most fine-grained sub-problem can be denoted as finding the optimal decomposition decision for the first $1$ layer of this network, which has $0$ optional decomposition position and $2$ compulsory decomposition positions after layer $0$ and layer $1$.

Similarly, we define the most fine-grained sub-problem in the backward propagation as finding the optimal decomposition decision for the last $1$ layer of this network, which has $0$ optional decomposition position and $2$ compulsory decomposition positions after layer $L+1$ and layer $L$.

\begin{algorithm}[htbp]
\setcounter{algorithm}{2}
 \caption{DP-based forward scheduling algorithm}
 \label{alg:algorithm3}
\textbf{Input}: Parameter Transmission Cost $\vec{pt_i}$, Forward Computation Cost $\vec{fc_i}$, Introduced Overhead $\Delta t$, Layer Number $L$\\
\textbf{Output}: Forward Decision $\vec{p}$
 \begin{algorithmic}[1]
  \FOR {$m$ in [0, $L$]}
    \FOR {$n$ in [0, $L$]}
    \STATE $F[m][n]$ $\leftarrow +\infty$ \hfill\COMMENT{initialization}
    \STATE $Path[m][n]$ $\leftarrow -1$  \hfill\COMMENT{initialize state transition path}
    \ENDFOR
  \ENDFOR
  \STATE $F[0][0]$ $\leftarrow 0$   \hfill\COMMENT{initial state}
   \FOR {$m$ in [1, $L$]}
    \FOR {$n$ in [1, $m$]}
     \FOR {$k$ in [0, $m-1$]}
        \STATE $T_{lst}$ $\leftarrow$ $\max{(F[k][n-1], n\cdot\Delta t+\sum_{1 \leq l \leq m}pt_i^l)}$
        \IF {$F[m][n] \leq T_{lst} + \sum_{k+1 \leq l \leq m}fc_i^l$}
            \STATE continue    \hfill\COMMENT{$T_{lst}$ stands for legal starting time}
        \ELSE
            \STATE $F[m][n] \leftarrow T_{lst} + \sum_{k+1 \leq l \leq m}fc_i^l$
            \STATE $Path[m][n] \leftarrow k$
        \ENDIF
    \ENDFOR
    \ENDFOR
    \ENDFOR
 \STATE $T_{forward} \leftarrow +\infty$ \hfill\COMMENT{initialize minimum forward cost}
 \STATE $trbk_s \leftarrow -1$ \hfill\COMMENT{initialize trace back steps}
 \FOR {$k$ in [1, $L$]}
 \IF {$T_{forward} \leq F[L][k]$}
 \STATE continue
 \ELSE
 \STATE $T_{forward} \leftarrow F[L][k]$ \hfill\COMMENT{get minimum}
 \STATE $trbk_s \leftarrow k$ \hfill\COMMENT{derive traceback steps}
 \ENDIF
 \ENDFOR
 \STATE $\vec{p} \leftarrow \vec{0}$, $trbk_{cl} \leftarrow L$ \hfill\COMMENT{current layer during traceback}
 \FOR {$k$ in [1,$trbk_s$]}
 \STATE $tmp \leftarrow Path[trbk_{cl}][trbk_s-k+1]$
 \IF {$1 \leq tmp \leq L-1$}
 \STATE $p_{tmp} \leftarrow 1$ \hfill\COMMENT{enable this decomposition position}
 \ENDIF
 \STATE $trbk_{cl} \leftarrow tmp$
 \ENDFOR
\RETURN $\vec{p}$
 \end{algorithmic}
 \end{algorithm}

\subsubsection{DP-based Scheduling Algorithms}
Firstly, we present the Bellman Equation (i.e., the state transition equation) of our forward scheduling algorithm, which demonstrates the relation between the cost of the larger problem and the cost of the sub-problems as follows:

\begin{small}
\begin{equation}
F[m][n]=
\begin{cases}
\min\limits_{0 \leq k < m}{\{\max{(F[k][n-1],}}\\ \ \ \ \ \ \ { n\cdot\Delta t}+\sum_{1 \leq l \leq m}pt_i^l) \\ \ \ \ \ \ \ + \sum_{k+1 \leq l \leq m}fc_i^l\} & \text{$ 1 \leq n \leq m \leq L$}\\\\
+\infty & \text{$0 \leq m<n \leq L$}\\
+\infty & \text{$m \ne 0, \ n=0$}\\
0 & \text{$m=0 , \ n=0$}
\end{cases}
\label{eq8}
\end{equation}
\end{small}where $m$ stands for the sub-problem for the first $m$ layers while $n$ indicates that there exist $n$ enabled decomposition positions, including the compulsory starting position after layer $0$. Note that the compulsory end position after layer $L$ is excluded for it does not invoke a new transmission mini-procedure. Moreover, the costs of the boundaries are set to $+\infty$.

This proposed DP-based scheduling algorithm for the forward propagation is detailed in Algorithm 3. To be noticed, $F[m][n]$ only records the minimum cost of the first $m$ layers with $n$ enabled decomposition positions, it is not aware of which positions have been selected. To trace back the selections, path information is required. As is shown in Algorithm 3 line 16, $Path[m][n]$ records the route where $F[m][n]$ achieves the minimum from. Note that the subscripts of $F[m][n]$ and $Path[m][n]$ start from $0$ where row $0$ and column $0$ are the boundaries.

Similarly, the Bellman Equation of the scheduling algorithm for the backward propagation is given as follows: 
\begin{small}
\begin{equation}
B[m][n]=
\begin{cases}
\min\limits_{0 \leq k < m}{\{\max{(B[k][n-1],}}\\ \ \ \ \ \ \sum_{L - m +1 \leq l \leq L}bc_i^l) + \Delta t \\ \ \ \ \ \ \ + \sum_{L-m + 1 \leq l \leq L-k}gt_i^l\} & \text{$ 1 \leq n \leq m \leq L$}\\\\
+\infty & \text{$0 \leq m<n \leq L$}\\
+\infty & \text{$m \ne 0, \ n=0$}\\
0 & \text{$m=0 , \ n=0$}
\end{cases}
\label{eq9}
\end{equation}
\end{small}where $m$ stands for the sub-problem for the last $m$ layers while $n$ indicates that there exist $n$ enabled decomposition positions, including the compulsory starting position after layer $L+1$. And the costs of the boundaries are set to $+\infty$ as well.

\begin{algorithm}[htbp]
\setcounter{algorithm}{3}
 \caption{DP-based backward scheduling algorithm}
 \label{alg:algorithm4}
\textbf{Input}: Backward Computation Cost $\vec{bc_i}$, Gradient Transmission Cost $\vec{gt_i}$, Introduced Overhead $\Delta t$, Layer Number $L$\\
\textbf{Output}: Backward Decision $\vec{g}$
 \begin{algorithmic}[1]
  \FOR {$m$ in [0, $L$]}
    \FOR {$n$ in [0, $L$]}
    \STATE $B[m][n]$ $\leftarrow +\infty$ \hfill\COMMENT{initialization}
    \STATE $Path[m][n]$ $\leftarrow -1$  \hfill\COMMENT{initialize state transition path}
    \ENDFOR
  \ENDFOR
  \STATE $B[0][0]$ $\leftarrow 0$ \hfill\COMMENT{initial state}
   \FOR {$m$ in [1, $L$]}
    \FOR {$n$ in [1, $m$]}
     \FOR {$k$ in [0, $m-1$]}
          \STATE $T_{lst}$ $\leftarrow$ $\max{(B[k][n-1], \sum_{L - m +1 \leq l \leq L}bc_i^l)}$
          \IF {$B[m][n] \leq T_{lst} + \Delta t + \sum_{L-m + 1 \leq l \leq L-k}gt_i^l$}
              \STATE continue
          \ELSE
              \STATE $B[m][n] \leftarrow T_{lst} + \Delta t + \sum_{L-m + 1 \leq l \leq L-k}gt_i^l$
              \STATE $Path[m][n] \leftarrow k$
          \ENDIF
  \ENDFOR
  \ENDFOR
  \ENDFOR
 \STATE $T_{backward} \leftarrow +\infty$, $trbk_s \leftarrow -1$ \hfill\COMMENT{traceback steps}
 \FOR {$k$ in [1, $L$]}
 \IF {$T_{backward} \leq F[L][k]$}
 \STATE continue
 \ELSE
 \STATE $T_{backward} \leftarrow F[L][k]$, $trbk_s \leftarrow k$
 \ENDIF
 \ENDFOR
\STATE $\vec{g} \leftarrow \vec{0}$, $trbk_{cl} \leftarrow 1$ \hfill\COMMENT{current layer during traceback}
 \FOR {$k$ in [1,$trbk_s$]}
 \STATE $tmp \leftarrow Path[L - trbk_{cl}+1][trbk_s-k+1]$
 \IF {$1 \leq tmp \leq L-1$}
 \STATE $g_{tmp} \leftarrow 1$ \hfill\COMMENT{enable this decomposition position}
 \ENDIF
 \STATE $trbk_{cl} \leftarrow L - tmp + 1$
 \ENDFOR
\RETURN $\vec{g}$
 \end{algorithmic} 
 \end{algorithm}

The DP-based scheduling algorithm for the backward propagation is given in Algorithm 4. As the study subject of the sub-problems shifts from the first $m$ layers to the last $m$ layers in the backward propagation, the path array's subscripts should be converted accordingly as is shown in Algorithm 4 line 31 and line 35.

\subsubsection{Optimal Substructure Guarantee}

Generally, if a problem can be solved optimally by breaking it into sub-problems and then recursively finding the optimal solutions to the sub-problems, then it is said to have an optimal substructure. In this section, we take our forward scheduling algorithm as an example to demonstrate that the communication scheduling problem with our crafted sub-problem definition has the property of optimal substructure.

Without loss of generality, assume there are $L$ layers in this targeted CNN model. To find the optimal decomposition decision in the forward propagation for this $L$-layer network, we can divide this problem into $L$ sub-problems: $F[L][1], F[L][2], ..., F[L][L]$ where $F[L][n]$ indicates the minimum cost for this $L$-layer problem with $n$ enabled decomposition positions, including the compulsory starting position after layer $0$. Then the optimal solution with minimum cost can be determined by $\min(F[L][1], F[L][2], ..., F[L][L])$.

As for the sub-problem $F[L][n]$, for any $ 1 \le n \le L$, it can be solved by finding the optimal solutions for the following $L$ sub-problems: $F[0][n-1], F[1][n-1], ..., F[L-2][n-1], F[L-1][n-1]$, and then enabling a new decomposition position (i.e., the decomposition position after layer $k$ if the sub-problem is $F[k][n-1]$, $ 0 \le k < L$) and comparing the costs of these $L$ cases to achieve the minimum.

Without loss of generality, for any $ 1 \le n \le m \le L$, $F[m][n]$ can be solved by finding the optimal solutions for the following $m$ sub-problems: $F[0][n-1], F[1][n-1], ..., F[m-1][n-1]$, then enabling a new decomposition position (i.e., the decomposition position after layer $k$ if the sub-problem is $F[k][n-1]$, $ 0 \le k < m$) and comparing the costs of these $m$ cases to achieve the minimum, recursively.

In this way, we can solve the forward scheduling problem optimally from the bottom to the top. Following the same principle, the optimal substructure of the backward scheduling algorithm is guaranteed as well.

\subsubsection{Complexity Analysis}
The space complexity of Algorithm 3 and Algorithm 4 is $\mathcal{O}(L^2)$ since $F[m][n]$, $B[m][n]$, $Path[m][n]$ are two-dimensional arrays with respect to the depth of the targeted network $L$. Considering that the partial accumulations of $\vec{pt_i}$, $\vec{fc_i}$, $\vec{bc_i}$, $\vec{gt_i}$ can be preprocessed and stored in several one-dimensional arrays or two-dimensional arrays, all of the summation results can be derived in $\mathcal{O}(1)$. Therefore, the time complexity of these two algorithms is $\mathcal{O}(L^3)$ since the $for$ loop is nested three levels. Commonly, the scheduling overheads of these two algorithms are negligible since an iteration's computation and communication overheads are orders of magnitude larger.

\subsection{Minimizing Profiling and Scheduling Overheads}
Although real-time profiling and scheduling will provide performance improvements, these processes themselves also bring some extra computation overheads. To minimize these overheads, the profiling switch and the scheduling algorithms can be enabled once per epoch as default. If the profiling results are relatively stable for all iterations in each epoch, we can conduct the communication scheduling decision of the first iteration for all iterations in the same epoch. If we train a CNN with 8 edge devices on the CIFAR-10 dataset and the total batch size is set to 256, then this whole process will be conducted only once per 195 iterations on each device. When the training enters the next epoch, the communication scheduling decision will then adapt to the latest condition accordingly. Of course, it is optional to manually configure a small interval (e.g., once per iteration) to keep the profiling results up-to-date. After all, it doesn't matter as long as the performance benefits outweigh the extra overheads.

Apart from this, since the backward computation procedure precedes the gradient transmission procedure in the backward propagation, the computation process unit is relatively vacant once the last computation mini-procedure is completed. Therefore, an idle-event-trigger is implemented in DynaComm to launch the scheduling algorithm for the forward propagation of the next iteration in advance to hide the scheduling overhead. Also, the backward scheduler can be launched while the edge device is waiting for the first layer's parameters in the forward propagation of the next iteration.

\begin{figure*}[htbp]
\centering

\subfigure[VGG-19]{
\begin{minipage}[t]{0.24\linewidth}
\centering
\includegraphics[width=136px]{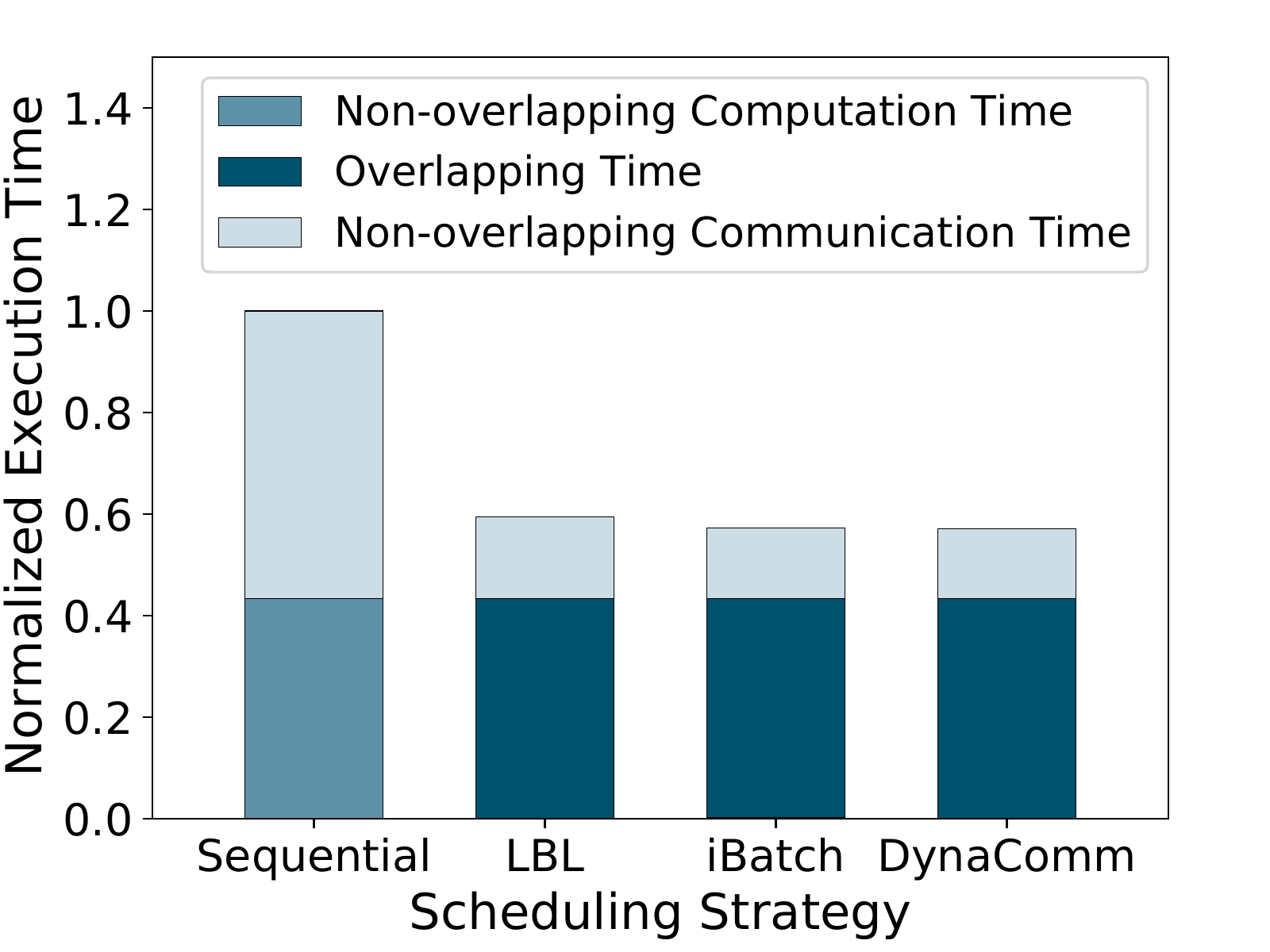}
%\caption{fig1}
\end{minipage}%
}%
\subfigure[GoogLeNet]{
\begin{minipage}[t]{0.24\linewidth}
\centering
\includegraphics[width=136px]{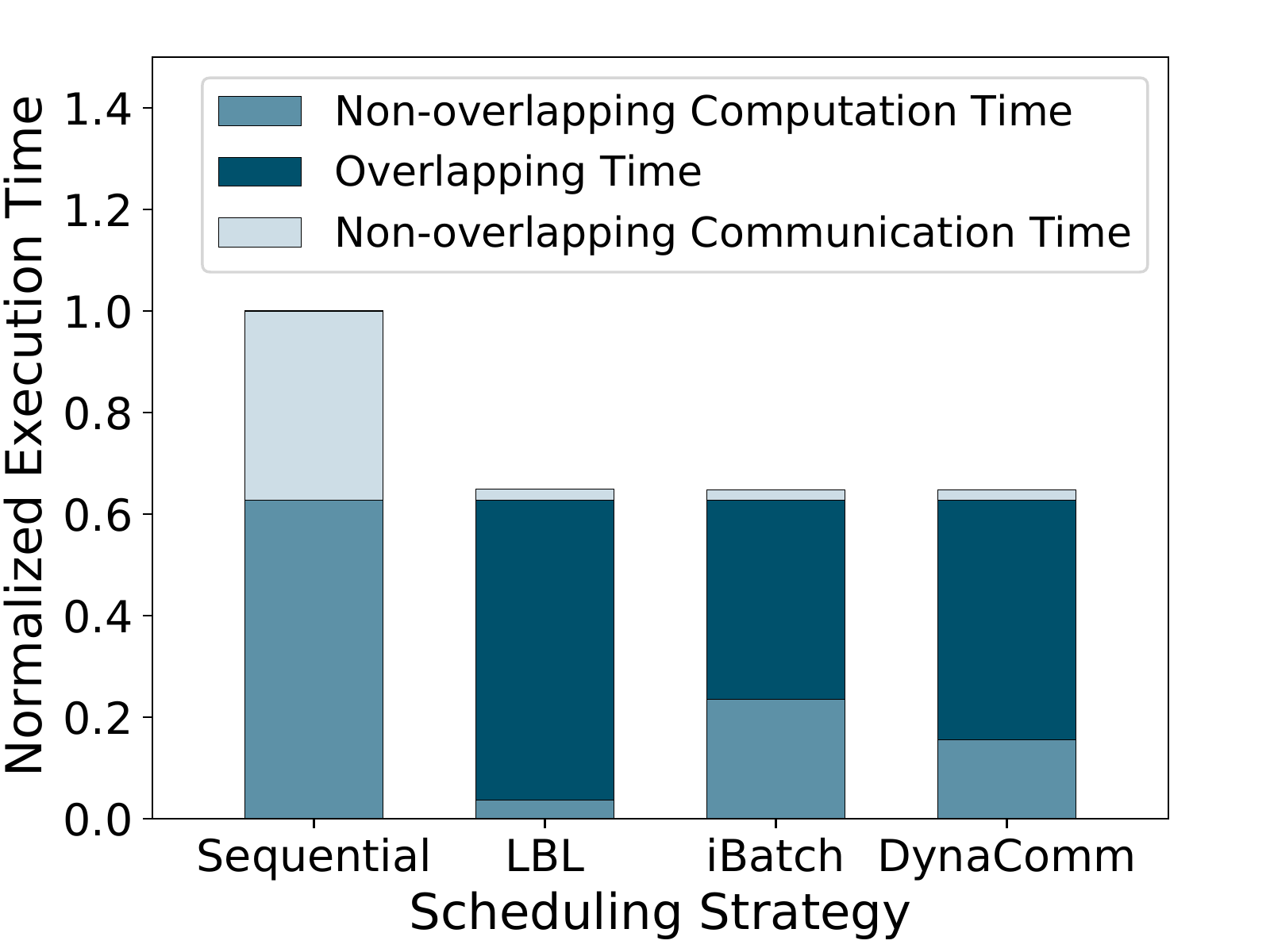}
%\caption{fig2}
\end{minipage}%
}%            
\subfigure[Inception-v4]{
\begin{minipage}[t]{0.24\linewidth}
\centering
\includegraphics[width=136px]{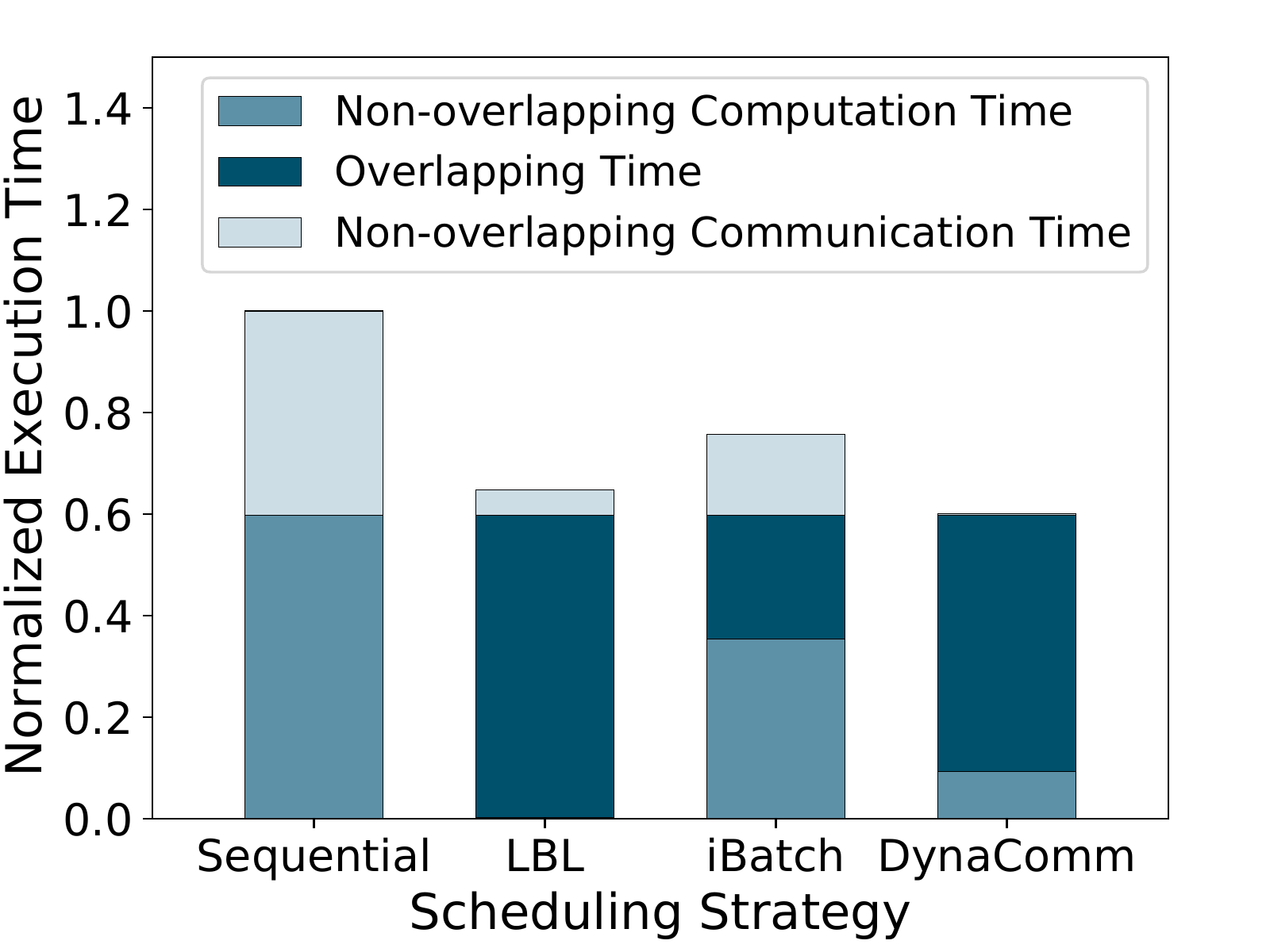}
%\caption{fig3}
\end{minipage}
}%
\subfigure[ResNet-152]{
\begin{minipage}[t]{0.24\linewidth}
\centering
\includegraphics[width=136px]{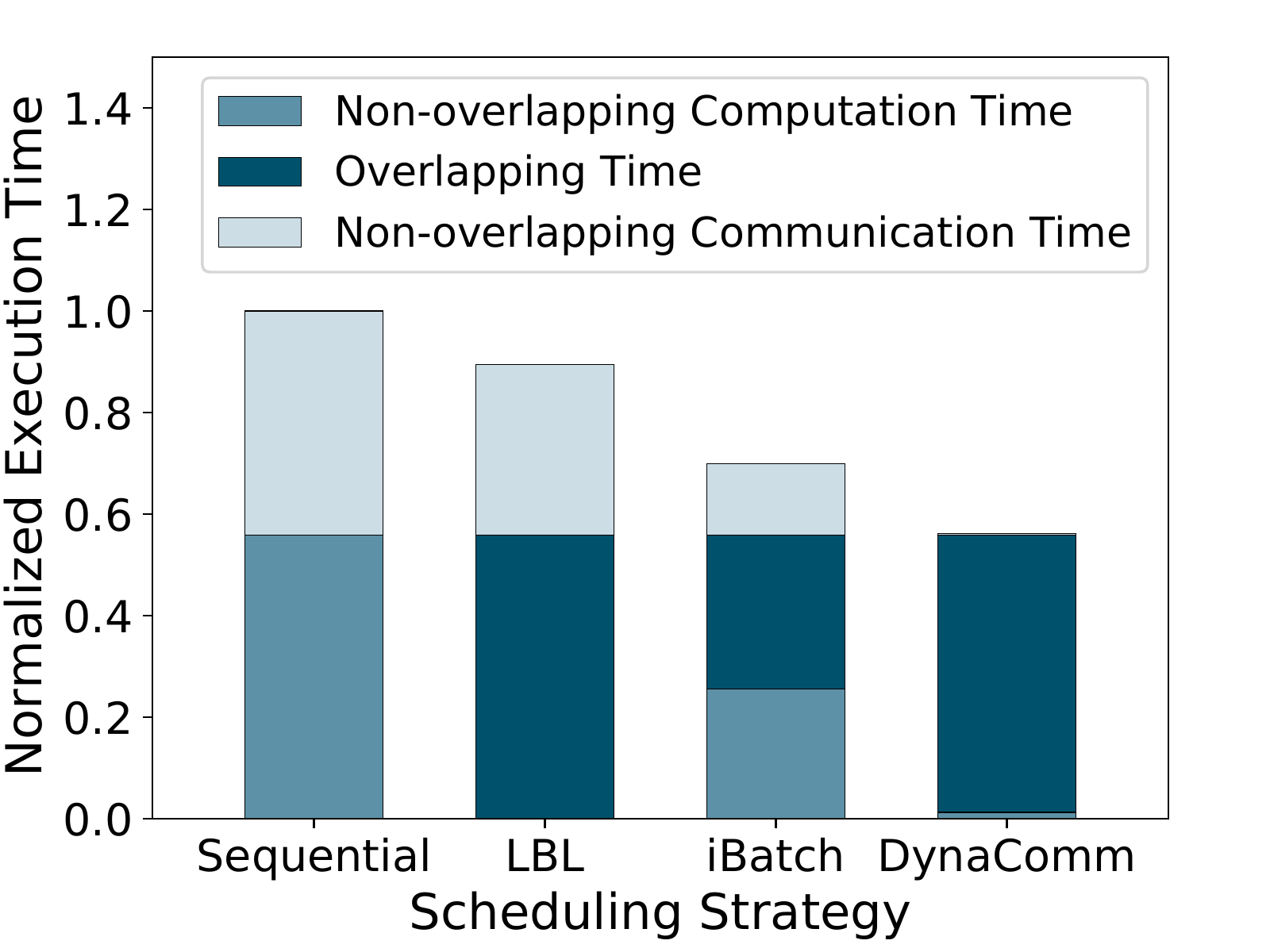}
%\caption{fig4}
\end{minipage}
}%

\centering
\caption{The normalized execution time of forward propagation when using different communication scheduling strategies (batch size = 32).}
\end{figure*}

\begin{figure*}[htbp]
\centering

\subfigure[VGG-19]{
\begin{minipage}[t]{0.24\linewidth}
\centering
\includegraphics[width=136px]{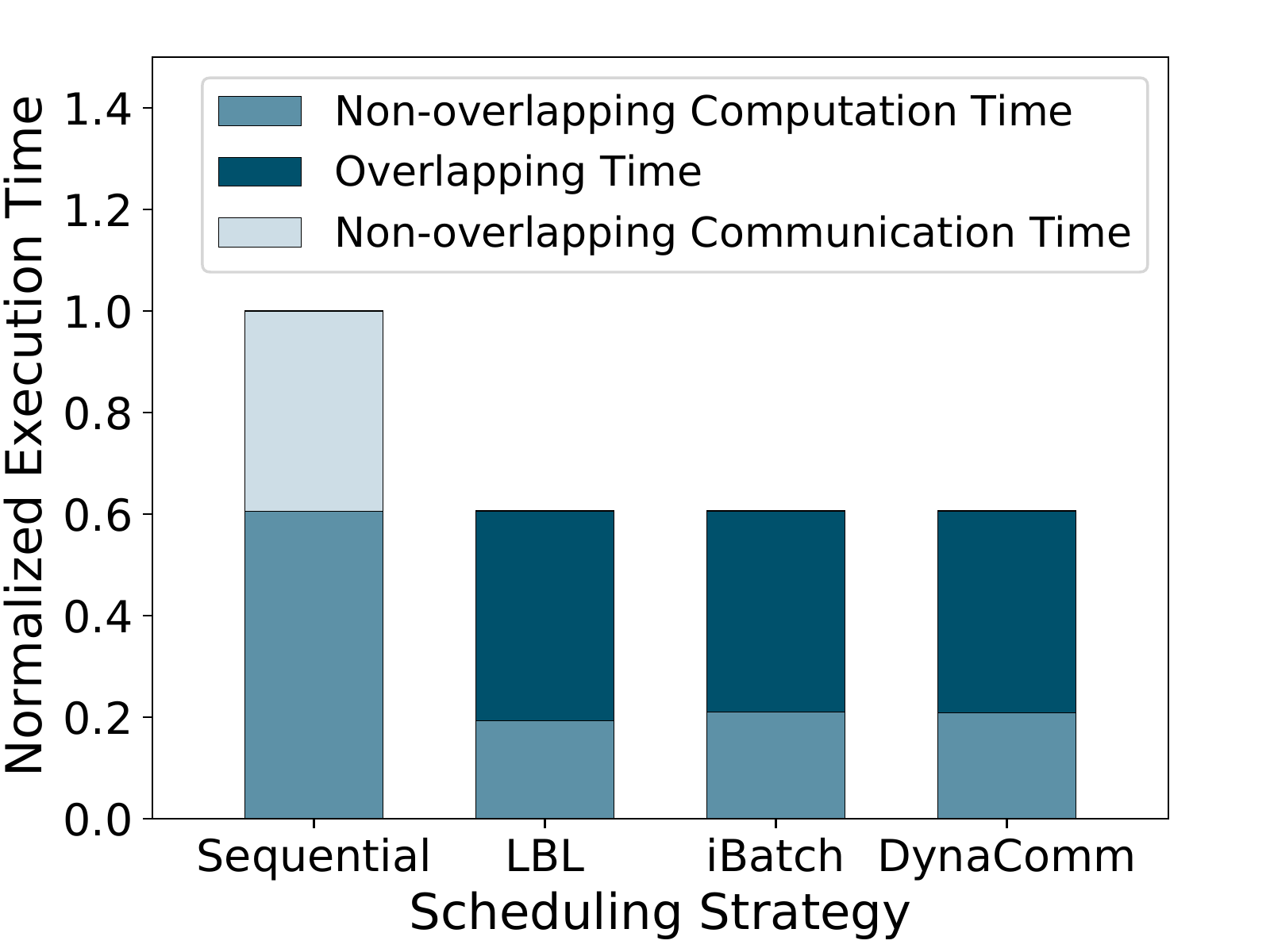}
%\caption{fig1}
\end{minipage}%
}%
\subfigure[GoogLeNet]{
\begin{minipage}[t]{0.24\linewidth}
\centering
\includegraphics[width=136px]{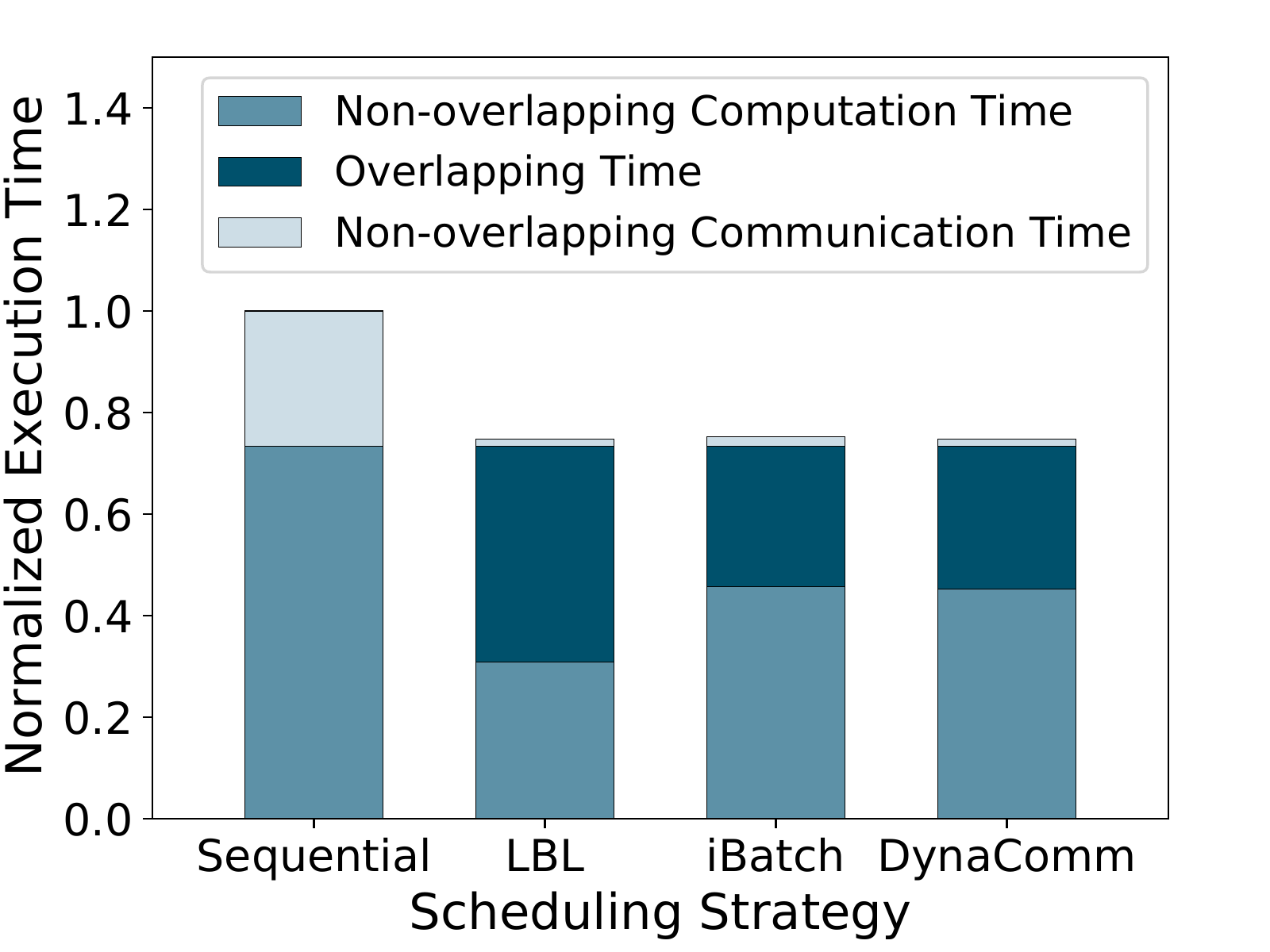}
%\caption{fig2}
\end{minipage}%
}%            
\subfigure[Inception-v4]{
\begin{minipage}[t]{0.24\linewidth}
\centering
\includegraphics[width=136px]{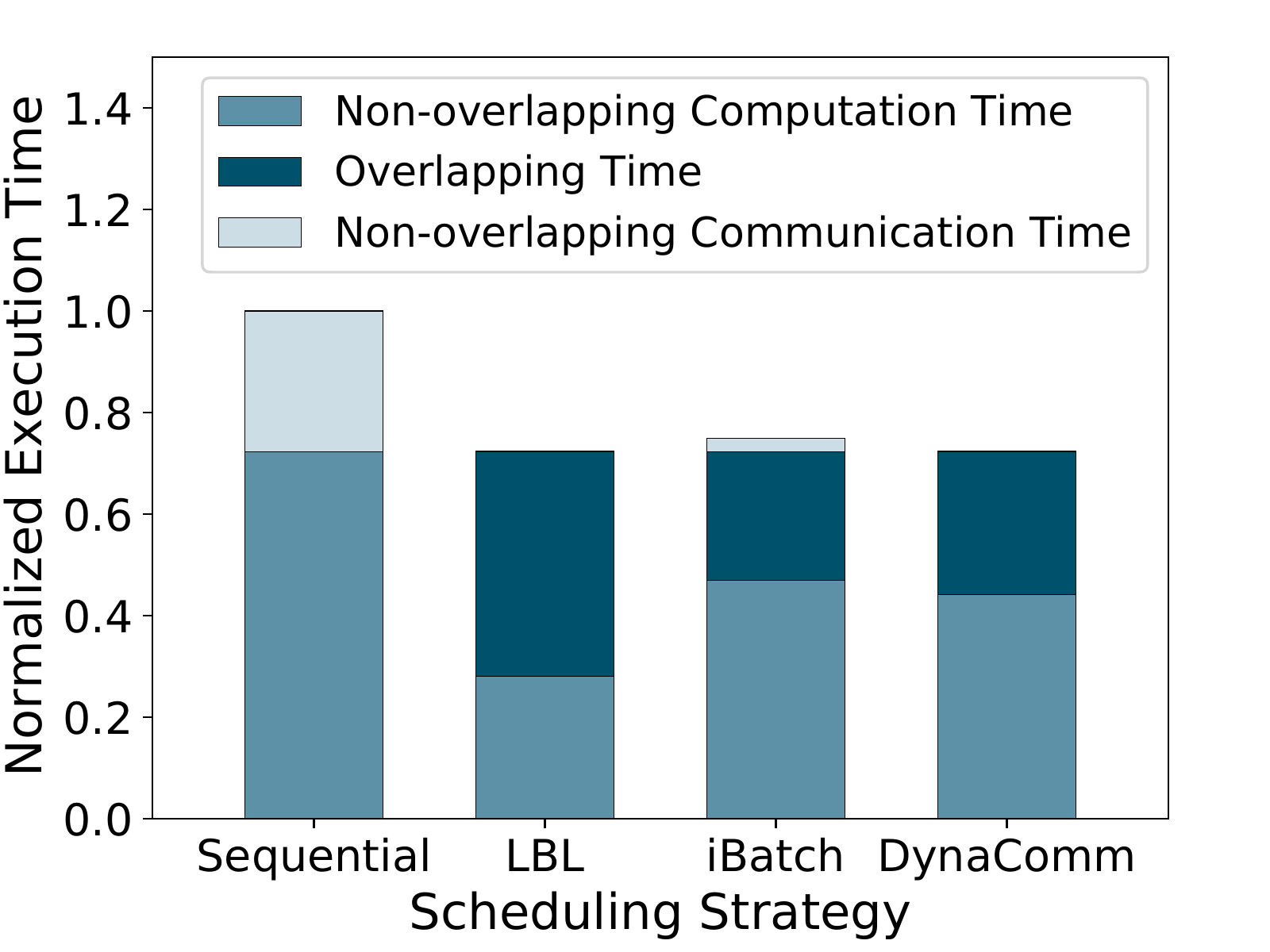}
%\caption{fig3}
\end{minipage}
}%
\subfigure[ResNet-152]{
\begin{minipage}[t]{0.24\linewidth}
\centering
\includegraphics[width=136px]{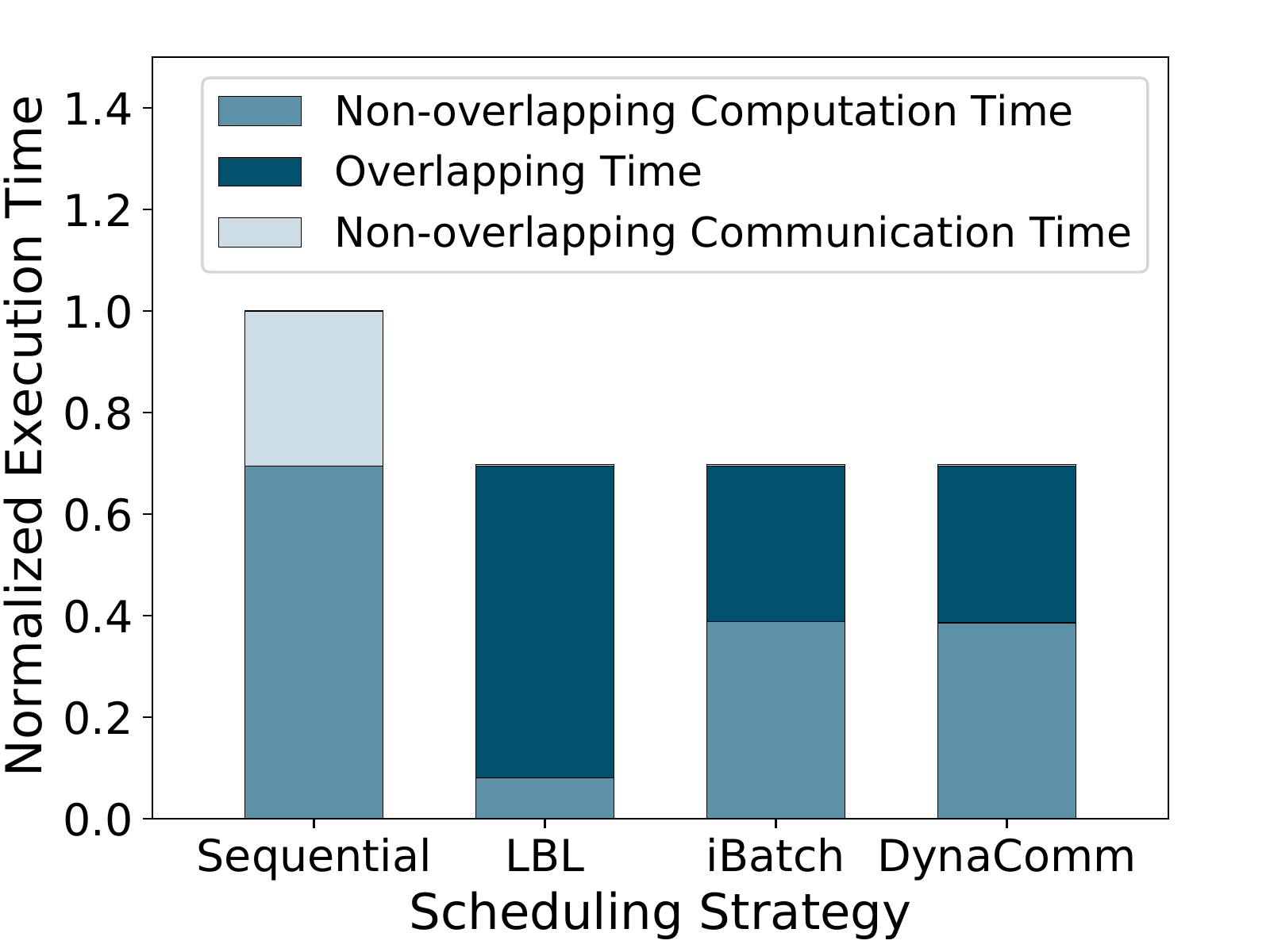}
%\caption{fig4}
\end{minipage}
}%

\centering
\caption{The normalized execution time of backward propagation when using different communication scheduling strategies (batch size = 32).}
\end{figure*}

\section{Performance Evaluation}
 
\subsection{Experimental Setup}
\subsubsection{Implementation and Testbed.}
Our experimental subject is a cluster of 8 machines (i.e., the edge servers in this study). Each machine is equipped with a 4-core Intel Xeon E3-1220 processor clocked at 3.00 GHz with 15717 MB of main memory. They both run Ubuntu 16.04.5 with a 3.10.0 Linux kernel. Although the RTT (i.e., Round-Trip Time) of the network between edges and clouds is hard to achieve a low number that is less than 1 ms (millisecond) as in a data center network, we believe that an RTT between 5 ms and 10 ms is achievable at present and feasible to provide a good user experience for edge computing applications. To imitate distributed CNN training over such network conditions, 4 parameter servers are deployed on a private cloud where the average RTT between the edge devices and the cloud is stable around 10 ms (i.e., min/avg/max/stddev = 8.280/10.337/11.434/1.023 ms). These parameter servers also run Ubuntu 16.04.5, and each one is provisioned with 16 GB of main memory and 4 vCPUs with up to 10 Gbps network bandwidth. As for the software layer, we implemented DynaComm and all the competing methods, which are the sequential execution scheme in the default PS (denoted as Sequential), the layer-by-layer transmission strategy (denoted as LBL), and iBatch (also known as iPart), in MXNet \cite{DBLP:journals/corr/ChenLLLWWXXZZ15} version 1.6.0 to conduct our experiments.

\subsubsection{Models and Dataset.} 
We focused on the computer vision applications that are broadly applied at the network edge. Therefore, we chose VGG-19 \cite{DBLP:journals/corr/SimonyanZ14a}, GoogLeNet \cite{DBLP:conf/cvpr/SzegedyLJSRAEVR15}, Inception-v4 \cite{DBLP:conf/aaai/SzegedyIVA17}, and ResNet-152 \cite{DBLP:conf/cvpr/HeZRS16} as our targeted models. Moreover, we chose ILSVRC12 \cite{DBLP:journals/ijcv/RussakovskyDSKS15} and CIFAR-10 \cite{krizhevsky2009learning} as the datasets in this study. All the models have been implemented in MXNet as built-in examples, and all model hyper-parameters remain default (except for batch size) during our experiments.

\subsubsection{Setups and Metrics.} 
To estimate the performance improvement brought by different layer-wise scheduling strategies and discuss their efficiencies, we conducted two sets of experiments with various networks for case studies, one with the batch size set to 32, the other with the batch size set to 16. Apart from this, we also varied batch size, bandwidth, and the number of workers to study the configuration sensitivity and the scalability of DynaComm. All these experiments are conducted on the ILSVRC12 dataset. We used the Normalized Execution Time, which is defined as the execution time over sequential implementation's overall execution time (i.e., the ratio of the optimized execution time to original implementation's total running time), the Iteration Time Reduced Ratio (i.e., the execution time reduced over sequential implementation's overall execution time), and Speedup (i.e., overall training speed over single worker training speed) to demonstrate and measure the effectiveness of DynaComm and its competing strategies. To verify that layer-wise scheduling does not violate the data dependency and the convergence of the training, we chose the training accuracy and validation accuracy as our measurements, and the experiments are performed on the CIFAR-10 dataset since this dataset is smaller and takes less time to train to convergence.

\subsection{Results and Analysis}

\textbf{Iteration Running Time Saving.}
In this section, we evaluate the iteration running time of the training process on four models (i.e., VGG-19 \cite{DBLP:journals/corr/SimonyanZ14a}, GoogLeNet \cite{DBLP:conf/cvpr/SzegedyLJSRAEVR15}, Inception-v4 \cite{DBLP:conf/aaai/SzegedyIVA17}, and ResNet-152 \cite{DBLP:conf/cvpr/HeZRS16}) with the ILSVRC12 \cite{DBLP:journals/ijcv/RussakovskyDSKS15} dataset as case studies. The iteration running time includes three portions, which are the Non-overlapping Computation Time, the Overlapping Time, and the Non-overlapping Communication Time. The Overlapping Time indicates that the communications and computations mini-procedures are executed in parallel during this period. Therefore, the whole iteration running time is reduced from the edge device's perspective. In the meanwhile, the Non-overlapping Time means that the communication or the computation mini-procedure is running alone and independently during that time due to data dependencies.

\begin{figure*}[htbp]
\centering

\subfigure[VGG-19]{
\begin{minipage}[t]{0.24\linewidth}
\centering
\includegraphics[width=136px]{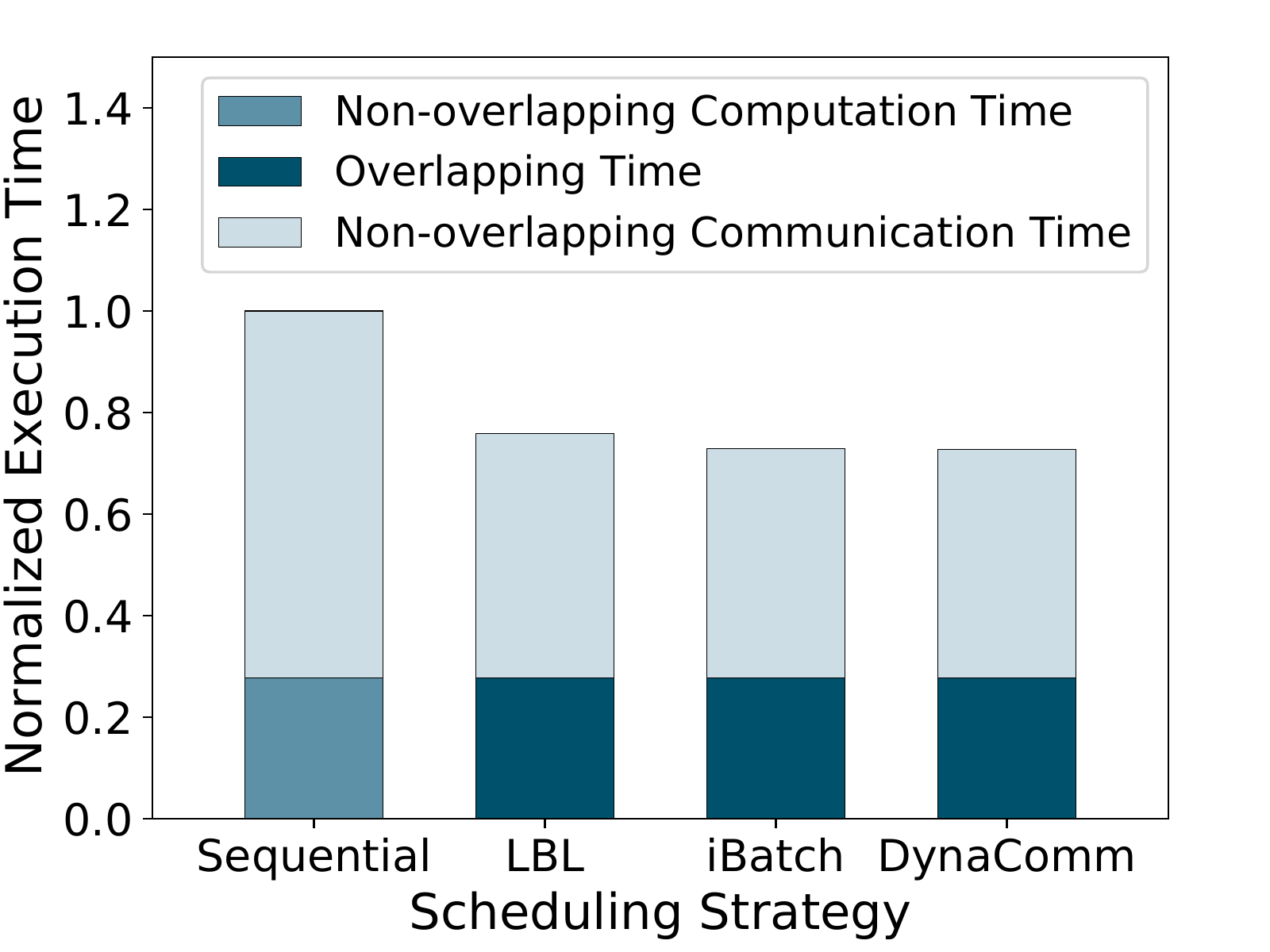}
%\caption{fig1}
\end{minipage}%
}%
\subfigure[GoogLeNet]{
\begin{minipage}[t]{0.24\linewidth}
\centering
\includegraphics[width=136px]{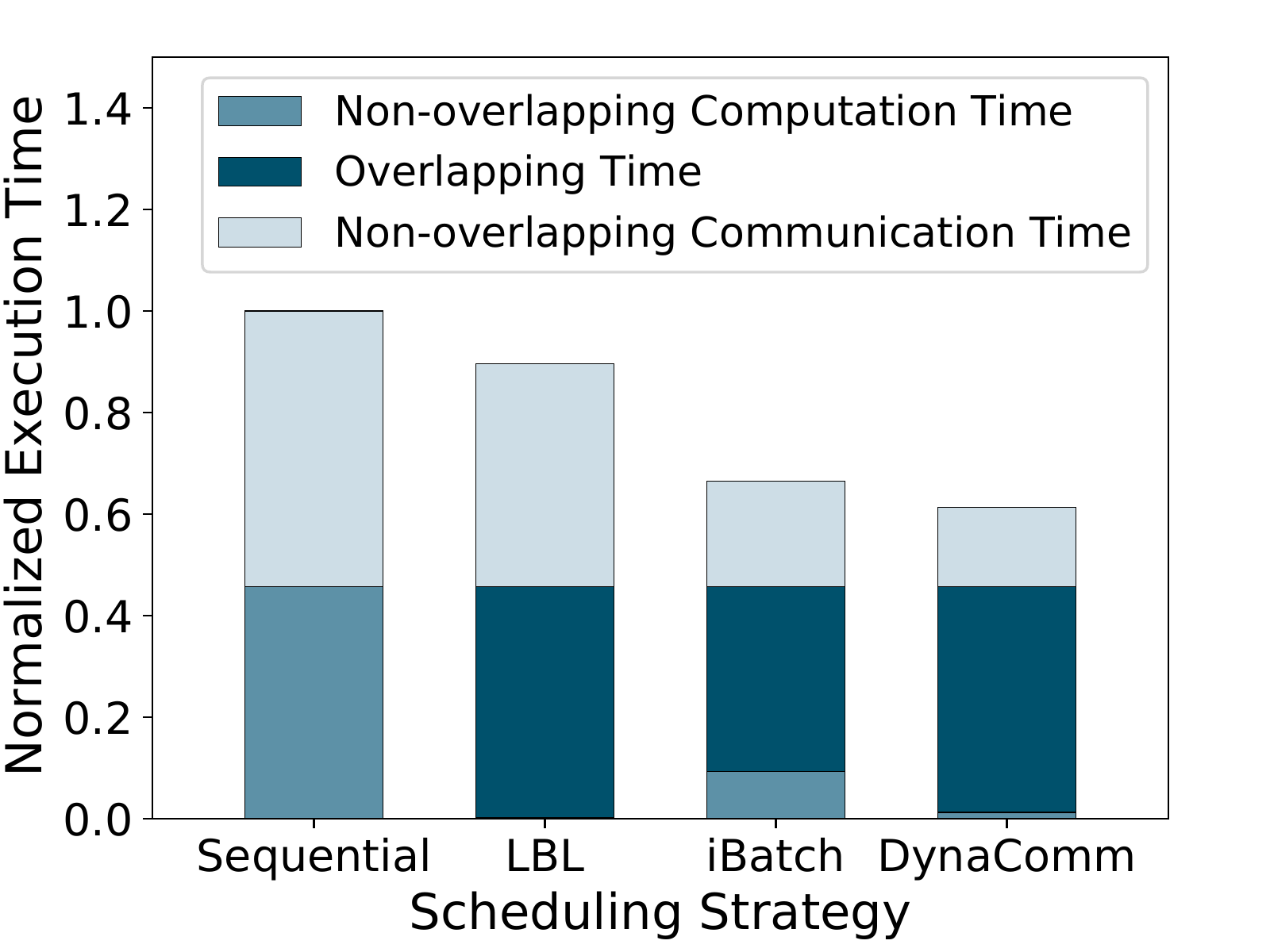}
%\caption{fig2}
\end{minipage}%
}%            
\subfigure[Inception-v4]{
\begin{minipage}[t]{0.24\linewidth}
\centering
\includegraphics[width=136px]{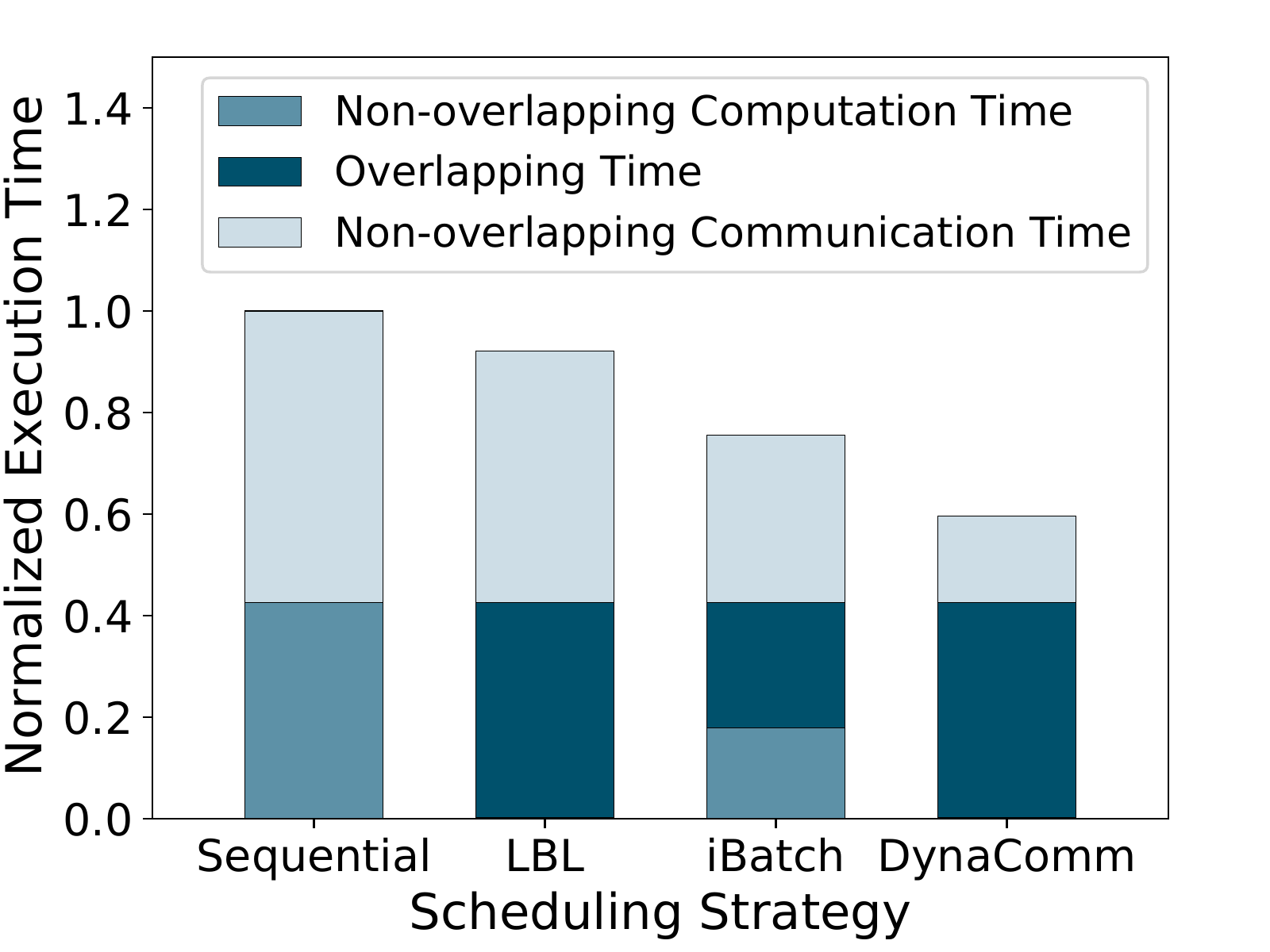}
%\caption{fig3}
\end{minipage}
}%
\subfigure[ResNet-152]{
\begin{minipage}[t]{0.24\linewidth}
\centering
\includegraphics[width=136px]{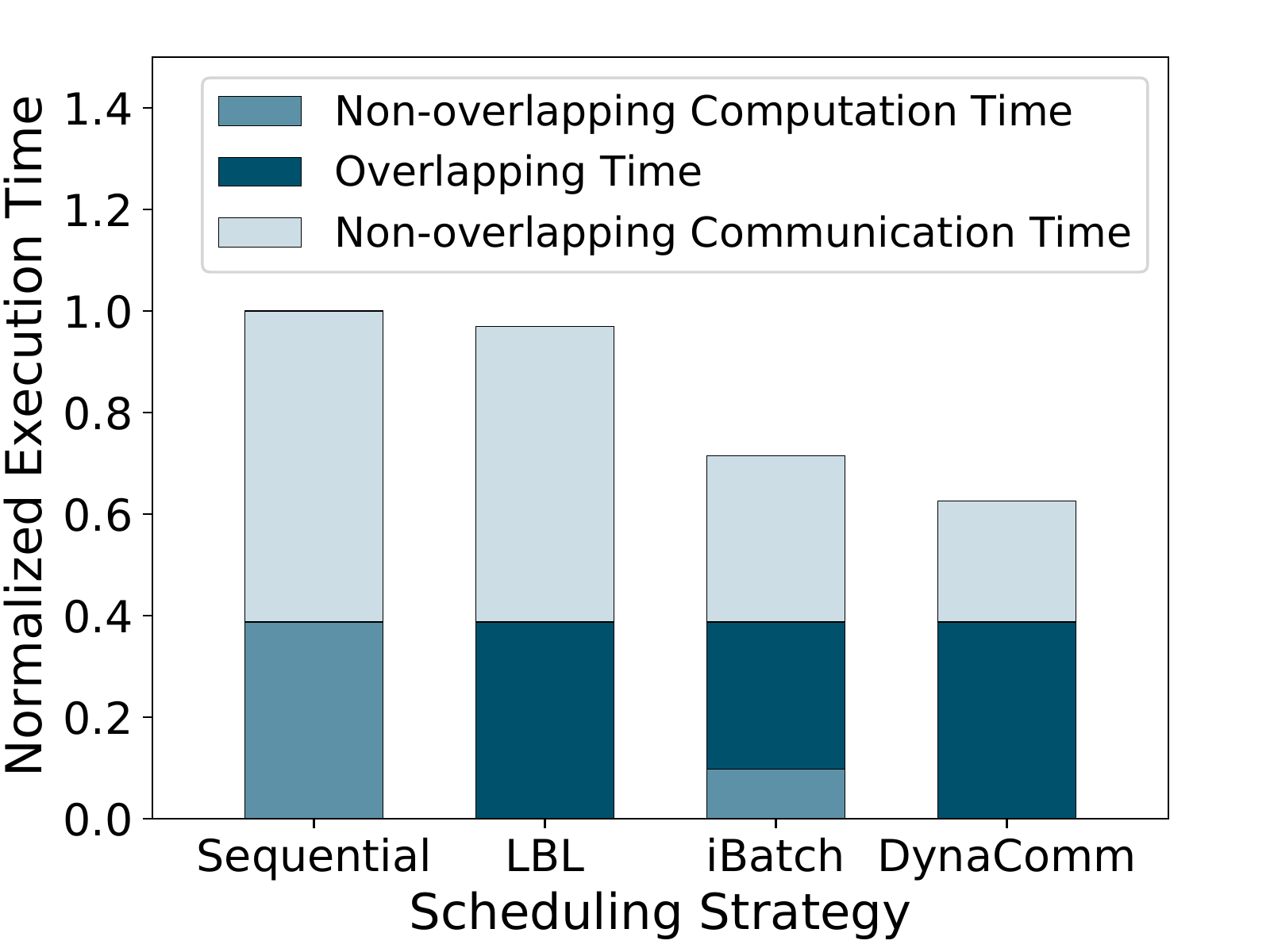}
%\caption{fig4}
\end{minipage}
}%

\centering
\caption{The normalized execution time of forward propagation when using different communication scheduling strategies (batch size = 16).}
\end{figure*}

\begin{figure*}[htbp]
\centering

\subfigure[VGG-19]{
\begin{minipage}[t]{0.24\linewidth}
\centering
\includegraphics[width=136px]{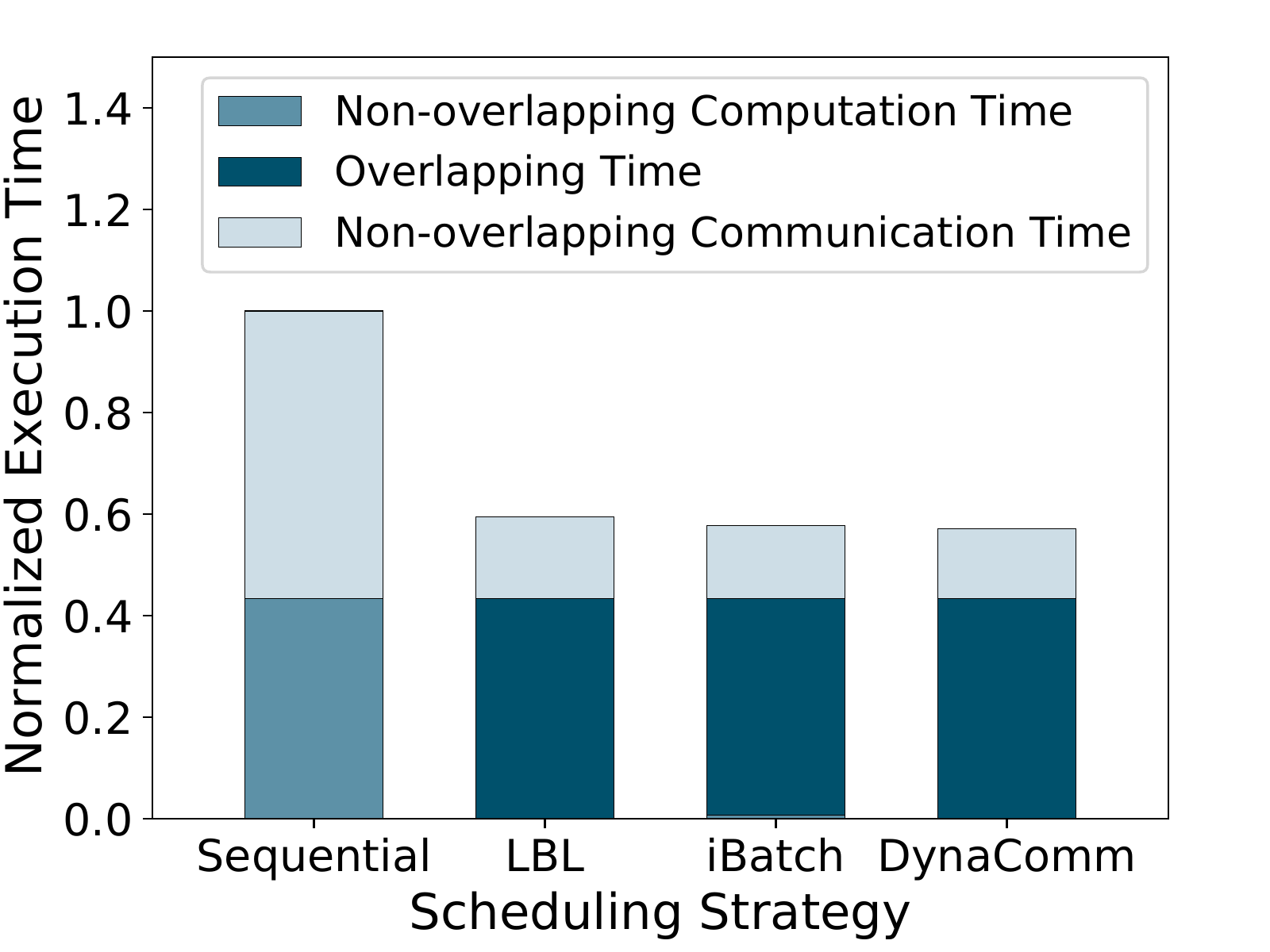}
%\caption{fig1}
\end{minipage}%
}%
\subfigure[GoogLeNet]{
\begin{minipage}[t]{0.24\linewidth}
\centering
\includegraphics[width=136px]{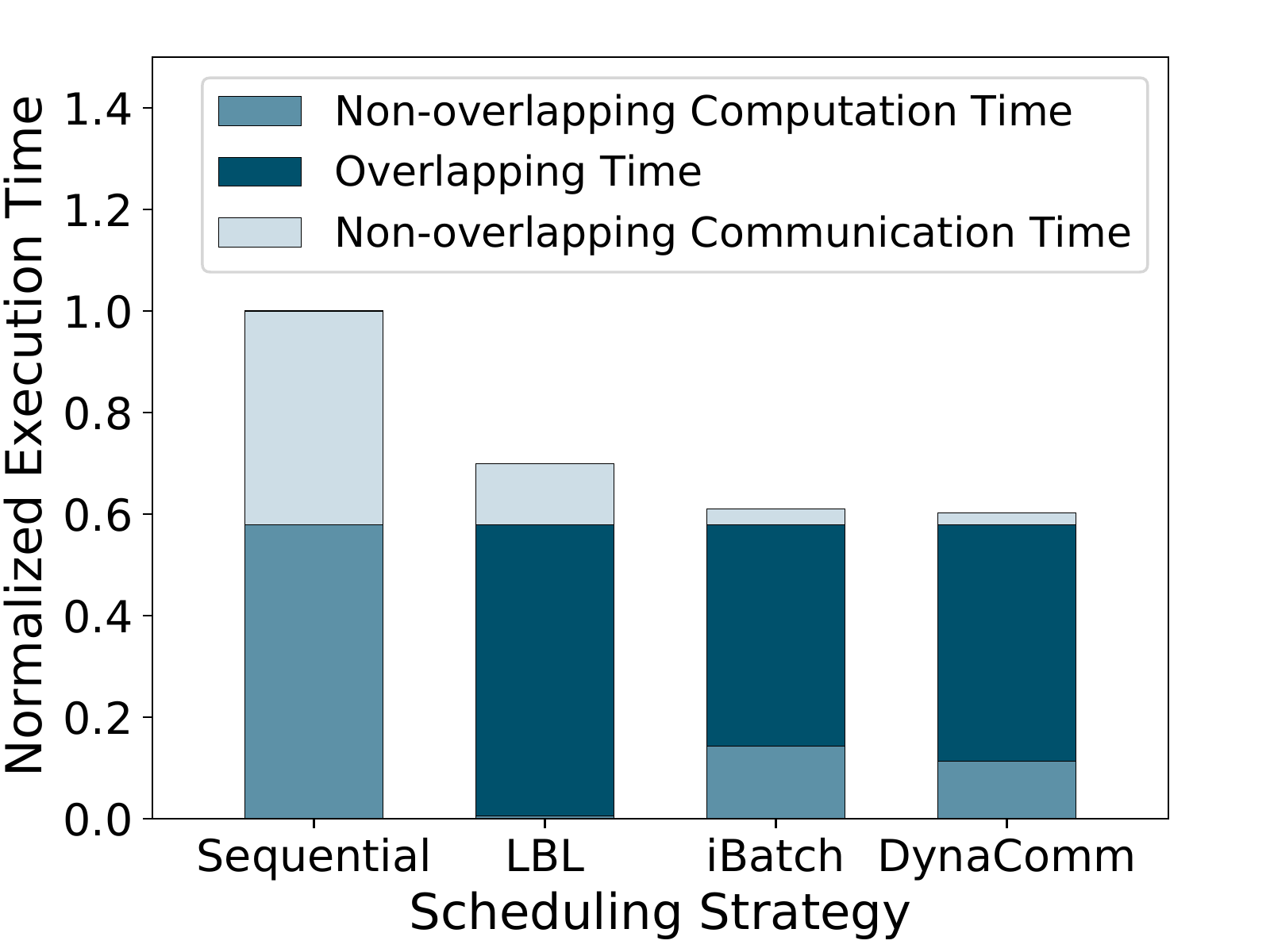}
%\caption{fig2}
\end{minipage}%
}%            
\subfigure[Inception-v4]{
\begin{minipage}[t]{0.24\linewidth}
\centering
\includegraphics[width=136px]{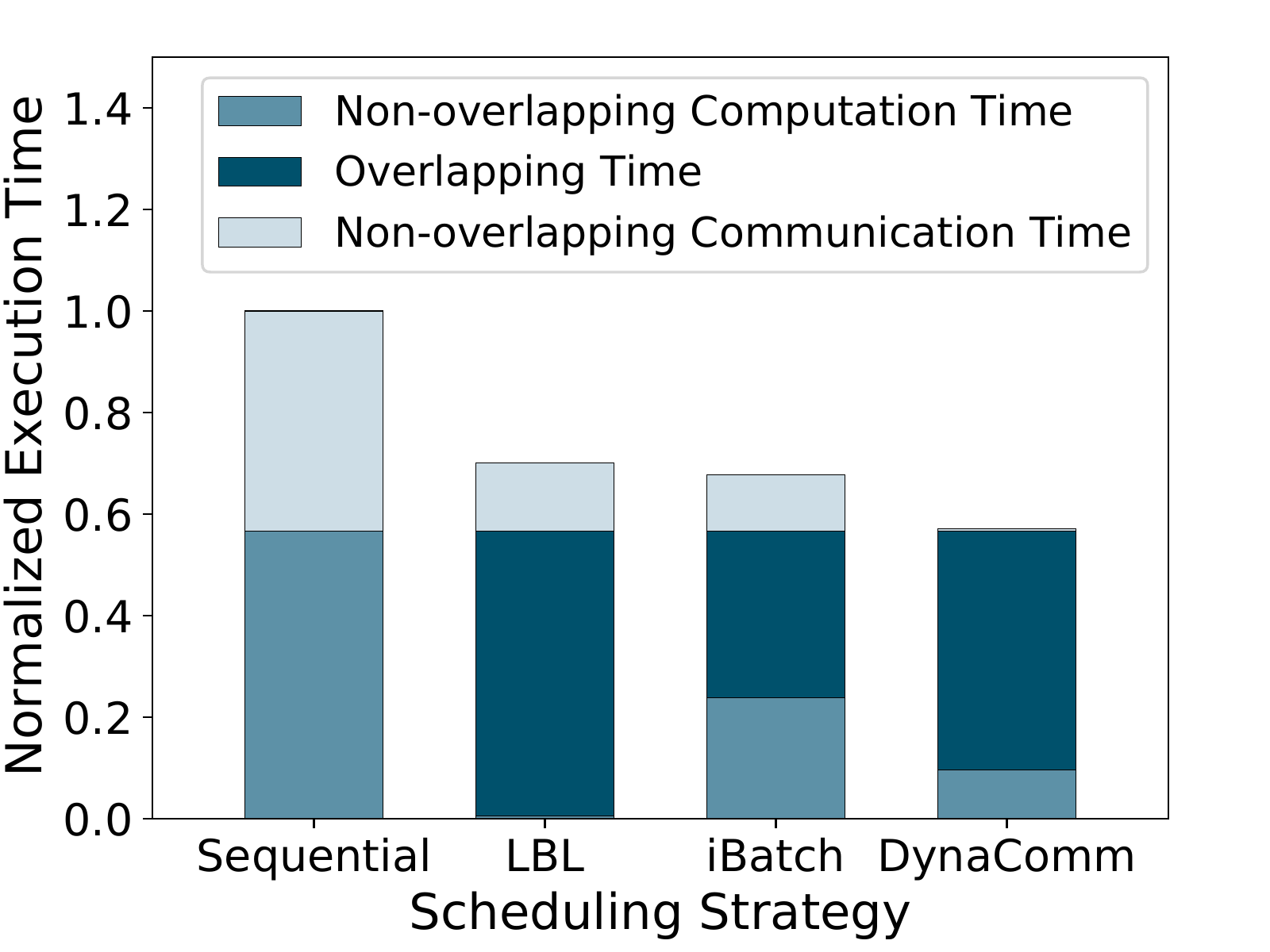}
%\caption{fig3}
\end{minipage}
}%
\subfigure[ResNet-152]{
\begin{minipage}[t]{0.24\linewidth}
\centering
\includegraphics[width=136px]{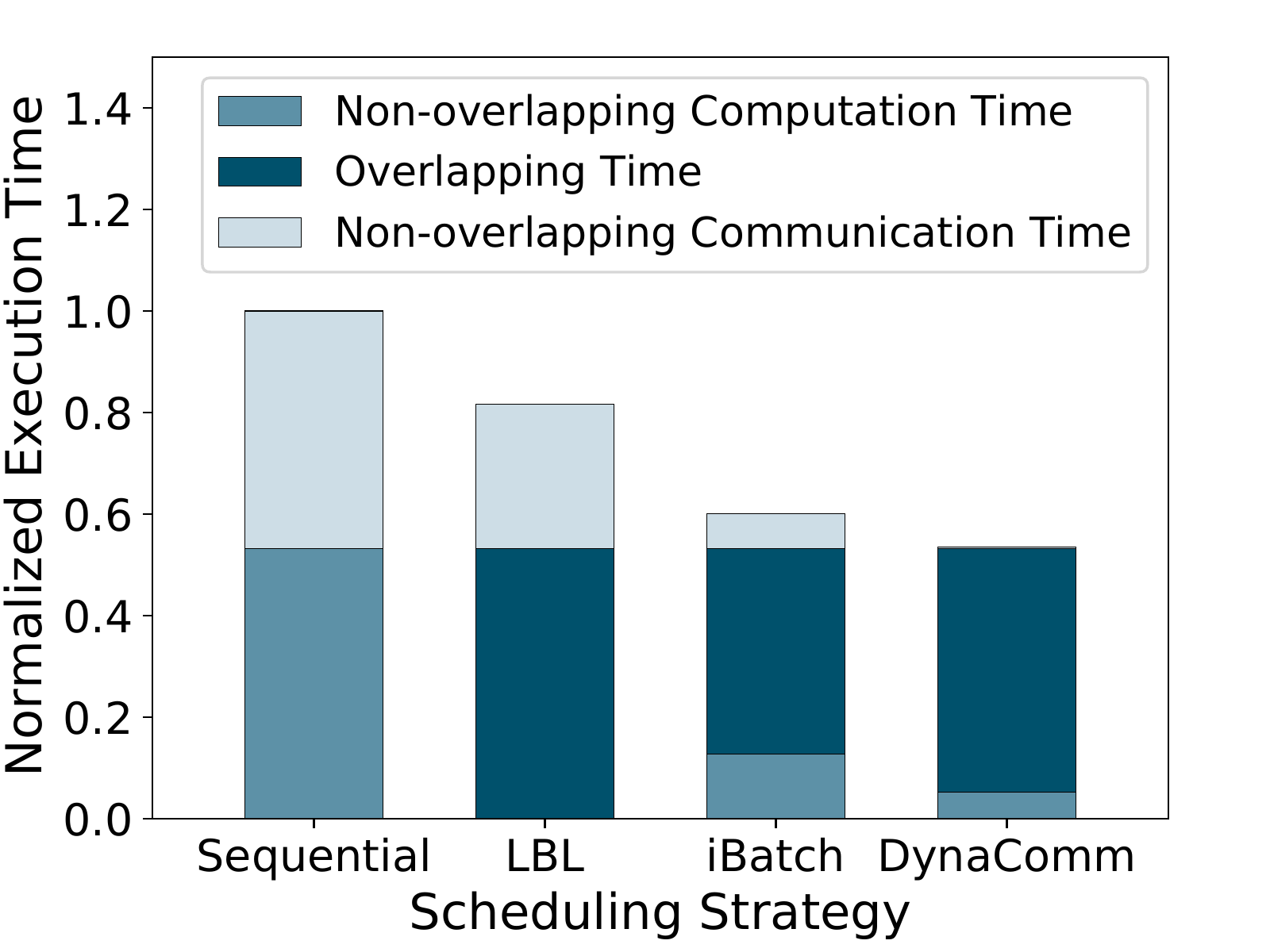}
%\caption{fig4}
\end{minipage}
}%

\centering
\caption{The normalized execution time of backward propagation when using different communication scheduling strategies (batch size = 16).}
\end{figure*}

Fig. 5 and Fig. 6 illustrate the normalized execution time when using different communication scheduling strategies to train various networks with the batch size set to 32. Similarly, Fig. 7 and Fig. 8 depict the normalized execution time with the batch size set to 16.

As is shown in Fig. 5 and Fig. 6, DynaComm manages to achieve optimal scheduling under all circumstances for both the forward and the backward propagation. When training on VGG-19, iBatch reduces the running time of the forward propagation by 42.79\% while DynaComm reduces the running time by 42.86\%. As for the backward propagation, they both reduce the running time by 39.35\%. Compared to these two strategies, the layer-by-layer transmission strategy also reduces the running time of the forward propagation by 40.54\% and the running time of the backward propagation by 39.35\%. The results imply that for the convolution neural networks that are not that deep such as VGG-19, all layer-wise strategies both achieve great and similar performance improvements. Therefore, the results of GoogLeNet are also similar to VGG-19. But the difference is that GoogLeNet is more computationally expensive while VGG-19's communication overhead dominates.

However, when the network goes deeper, iBatch and the layer-by-layer transmission strategy did not manage to achieve optimal scheduling due to their flaws. As is shown in Fig. 5 (c) and Fig. 5 (d), the layer-by-layer transmission strategy reduces the running time of the forward propagation on Inception-v4 and ResNet-152 by 35.25\% and 10.56\%, respectively. This is mostly because the layer-by-layer transmission strategy is a fixed solution, which means that it can not adapt to specific networks and run-time conditions. Therefore, when training ResNet-152, the layer-by-layer transmission strategy did not handle the transmission procedures of the fully connected layers very well, which takes up a lot of time in the final stage. As for iBatch, the numbers are 24.22\% and 30.02\%, respectively. The reason that iBatch performs poorly on Inception-v4 is that the greedy tactic applied commonly leads to a locally optimal solution when training deeper networks. Sometimes, the performance gain is acceptable, as is shown in Fig. 5 (d). But other times, it performs even worse than the vanilla layer-by-layer transmission strategy, which is exhibited in Fig. 5 (c). Such unstable and unpredictable performance may cause some problems when applied in real use cases.

Different from these two strategies, DynaComm reduces the running time of the forward propagation on Inception-v4 and ResNet-152 by 39.99\% and 43.84\%, respectively. DynaComm can adapt to any network and any run-time conditions for it assures that the scheduling decision always achieves the optimal guaranteed by the characteristic of optimal substructure mentioned in Section IV-B. In the meanwhile, the results for the backward propagation are 27.55\% and 30.29\%. Therefore, DynaComm manages to achieve optimal scheduling for all networks compared to competing strategies, and it reduces the total iteration running time on VGG-19, GoogLeNet, Inception-v4, and ResNet-152 by 41.10\%, 30.19\%, 33.78\%, and 37.06\%, respectively.

As for the backward propagation, DynaComm guarantees that almost all of the communication overheads are overlapped for all cases, as is shown in Fig. 6. However, the performance improvement in the backward propagation brought by DynaComm has reached the limit since the extra computation costs are inevitable. The vanilla layer-by-layer transmission strategy and iBatch both achieve close results. The only difference is that the vanilla layer-by-layer transmission strategy enables more gradient transmission mini-procedures than iBatch and DynaComm. Therefore, we can conclude that when the order of magnitude of computation dominates, the maximal performance gains from communication scheduling methods are depending on the proportion of communication.

From Fig. 7, we can observe that when the batch size is halved (i.e., the computation workloads are reduced), there are a certain amount of communication costs that can not be hidden in the forward propagation. Therefore, the proportion of computations can be regarded as the upper bound of the performance gains through communication scheduling methods. Although, DynaComm still manages to achieve optimal scheduling for all cases compared to competing strategies. Even though all computations are overlapped with communications when using the layer-by-layer transmission strategy, more enabled mini-procedures stand for more introduced coordination overheads. With the minimum introduced overheads, DynaComm reduces the running time of the forward propagation on VGG-19, GoogLeNet, Inception-v4, and ResNet-152 by 27.26\%, 38.64\%, 40.37\%, and 37.42\%, respectively.

\begin{figure}[t]
\centering

\subfigure[Iteration time reduced ratio versus batch size (bandwidth = 10 Gbps).]{
\begin{minipage}[t]{\linewidth}
\centering
\includegraphics[width=204px]{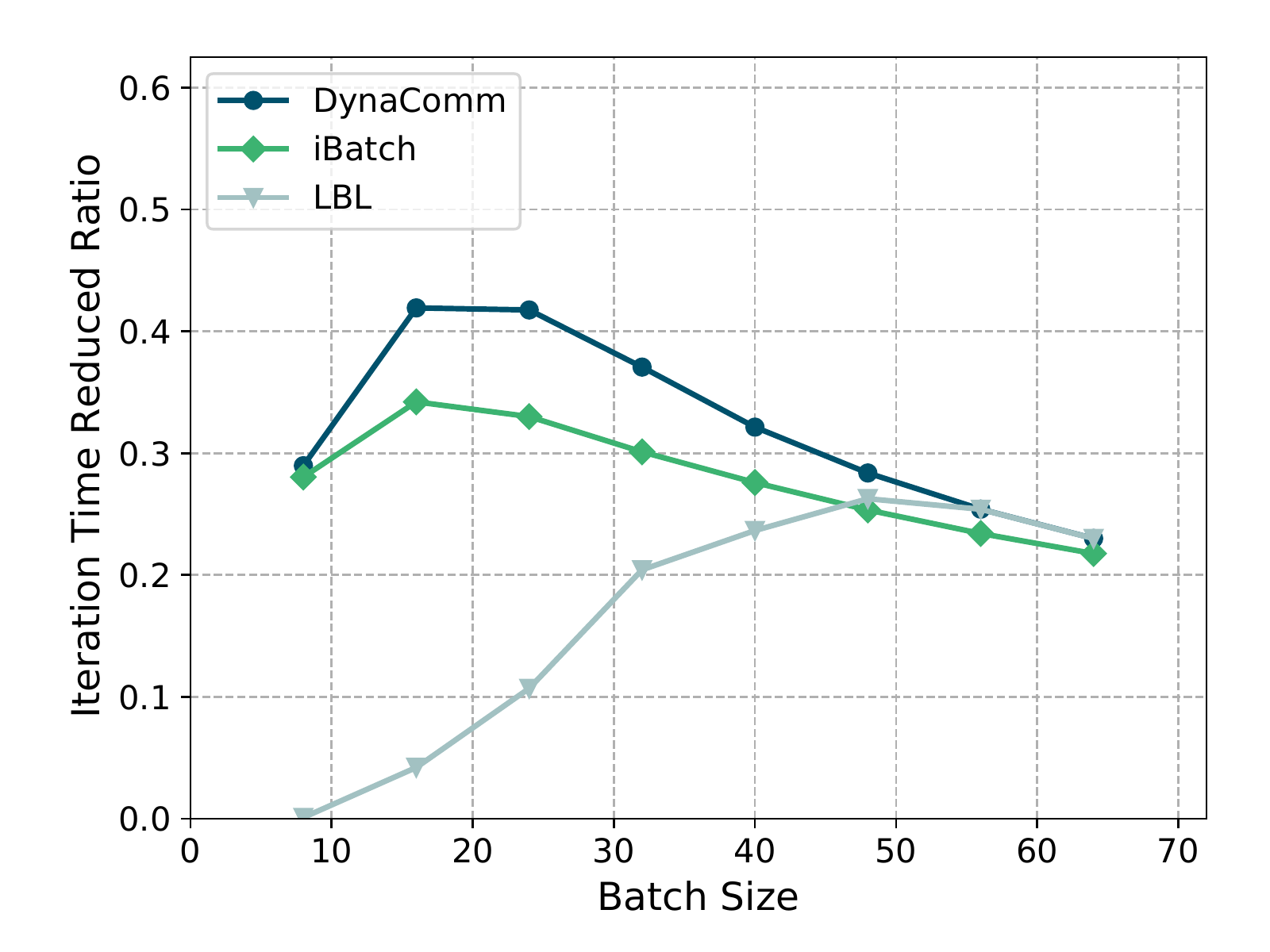}
%\caption{fig3}
\end{minipage}
}%

\subfigure[Iteration time reduced ratio versus bandwidth (batch size = 32)]{
\begin{minipage}[t]{\linewidth}
\centering
\includegraphics[width=204px]{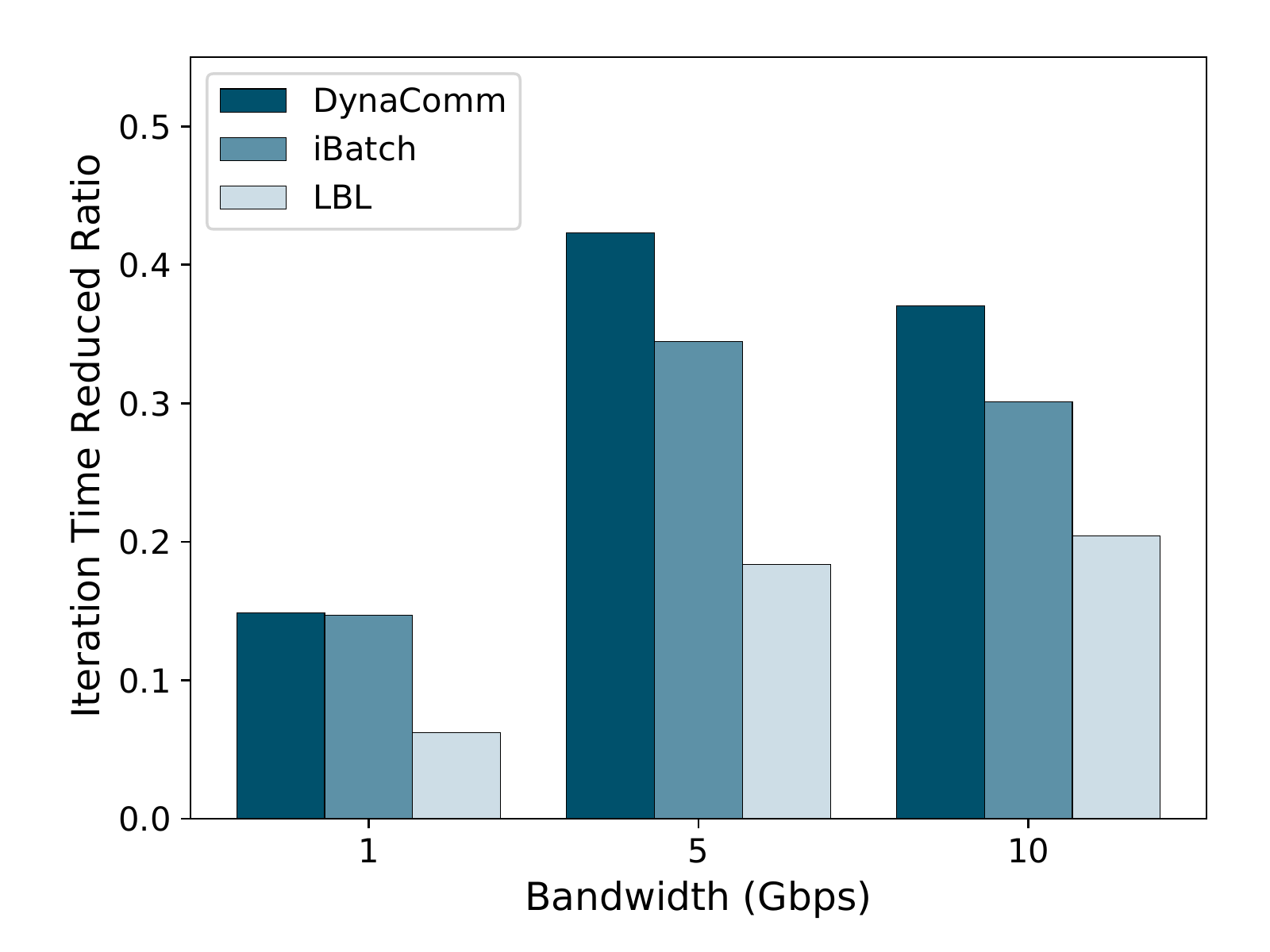}
%\caption{fig4}
\end{minipage}
}%

\centering
\caption{Iteration running time reduced ratio versus batch size and bandwidth with ResNet-152 on the ILSVRC12 dataset.}
\end{figure}

On the other hand, since the computations of the backward propagation are more time-consuming than the computations of the forward propagation, the computation/communication ratio is relatively balanced when the batch size is set to 16, as depicted in Fig. 8. For such cases, DynaComm manages to get a higher performance boost for the backward propagation. It reduces the running time of the backward propagation on VGG-19, GoogLeNet, Inception-v4, and ResNet-152 by 42.83\%, 39.80\%, 42.93\%, and 46.42\%, respectively. Due to the bigger performance gains in the backward propagation, the total iteration running time on VGG-19, GoogLeNet, Inception-v4, and ResNet-152 is reduced by 35.05\%, 39.22\%, 41.65\%, and 41.92\%, respectively. To achieve a higher performance improvement, the setting up of batch size has become a critical point, which requires consideration for both the forward and the backward propagation.

\textbf{Sensibility Analysis.} To investigate DyanComm's sensibility to the computation/communication ratio, we varied the batch size and the bandwidth (achieved through the Linux tc tool) in this subsection. All experiments run with ResNet-152 on the ILSVRC12 dataset. As is shown in Fig. 9 (a), both DynaComm and iBatch achieve higher performance improvement gains compared to the layer-by-layer transmission strategy when the batch size increases from a small value to a proper value at first. Yet when the batch size becomes greater than 24, the computation proportion starts to dominate and bottleneck the whole system, and the performance improvement brought by all scheduling methods is constrained by the computation proportion as mentioned before. Also, we can observe that iBatch fails to achieve optimal scheduling that it performs even worse than the layer-by-layer transmission strategy when the batch size grows greater than 48.

As for the bandwidth, all communication scheduling methods do not perform well when the bandwidth is 1 Gbps, as is shown in Fig. 9 (b). For this case, the communication is slow and congested, and the time spend on communication dominates and bottlenecks the whole system. Then when the bandwidth increases to 5Gbps, the performance gains brought by DynaComm and iBatch are significantly improved since the computation costs and communication costs are relatively balanced, which yields a bigger room for communication scheduling. When the bandwidth reaches 10 Gbps, the iteration time reduced ratio decreases for both DynaComm and iBatch. Although the overall iteration running time is reduced, the computation/communication ratio grows, and the computation proportion becomes the upper bound of the performance gains. In this case, decreasing the batch size (e.g., from 32 to 16) to balance the computations and communications can help improve the efficiency of the whole system.

\begin{figure}[t]
\centering
\subfigure[Top-1 training accuracy]{
\begin{minipage}[t]{0.475\linewidth}
\centering
\includegraphics[width=122px]{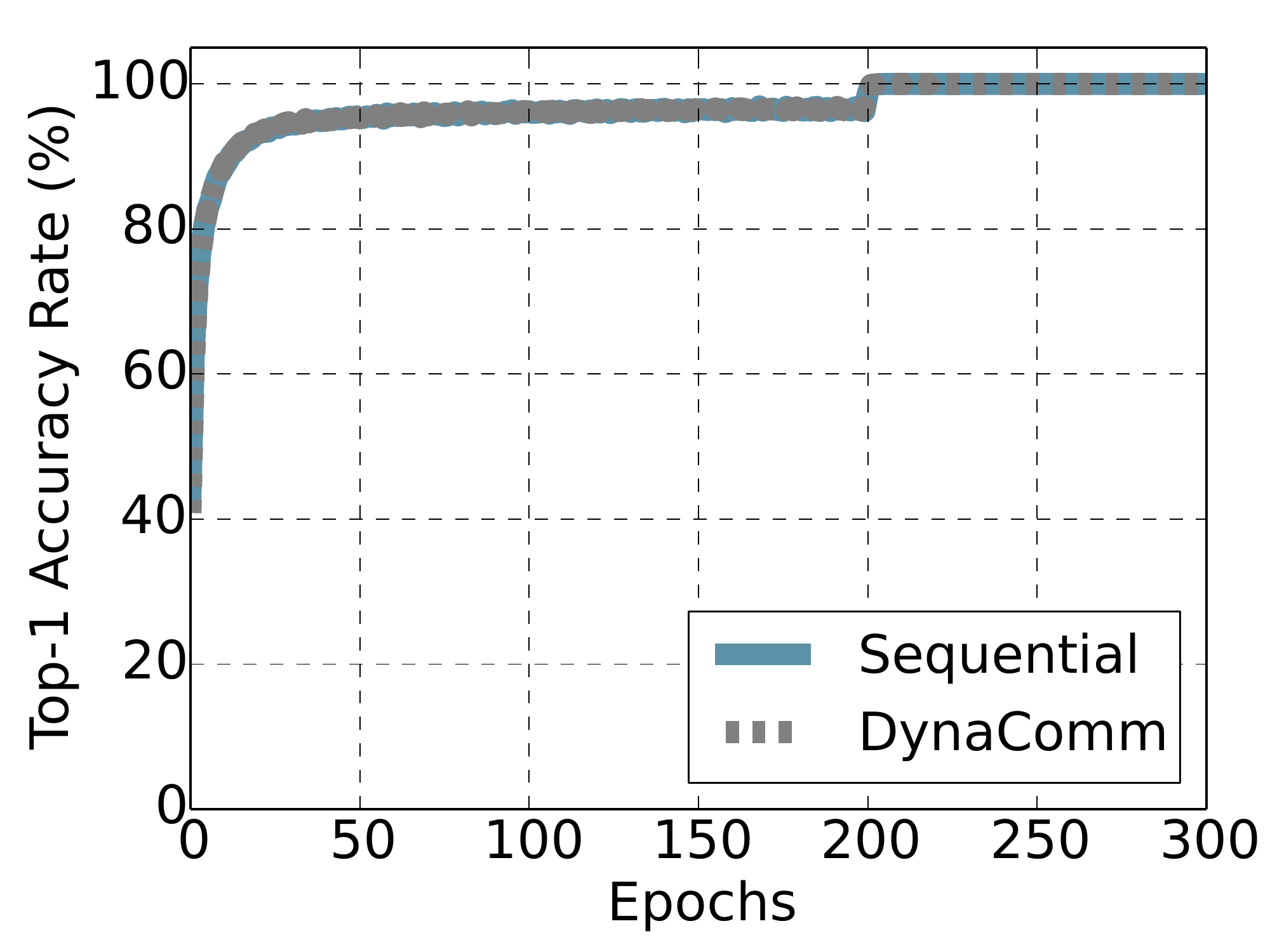}
%\caption{fig1}
\end{minipage}%
}%
\hfill
\subfigure[Top-5 training accuracy]{
\begin{minipage}[t]{0.4835\linewidth}
\centering
\includegraphics[width=122px]{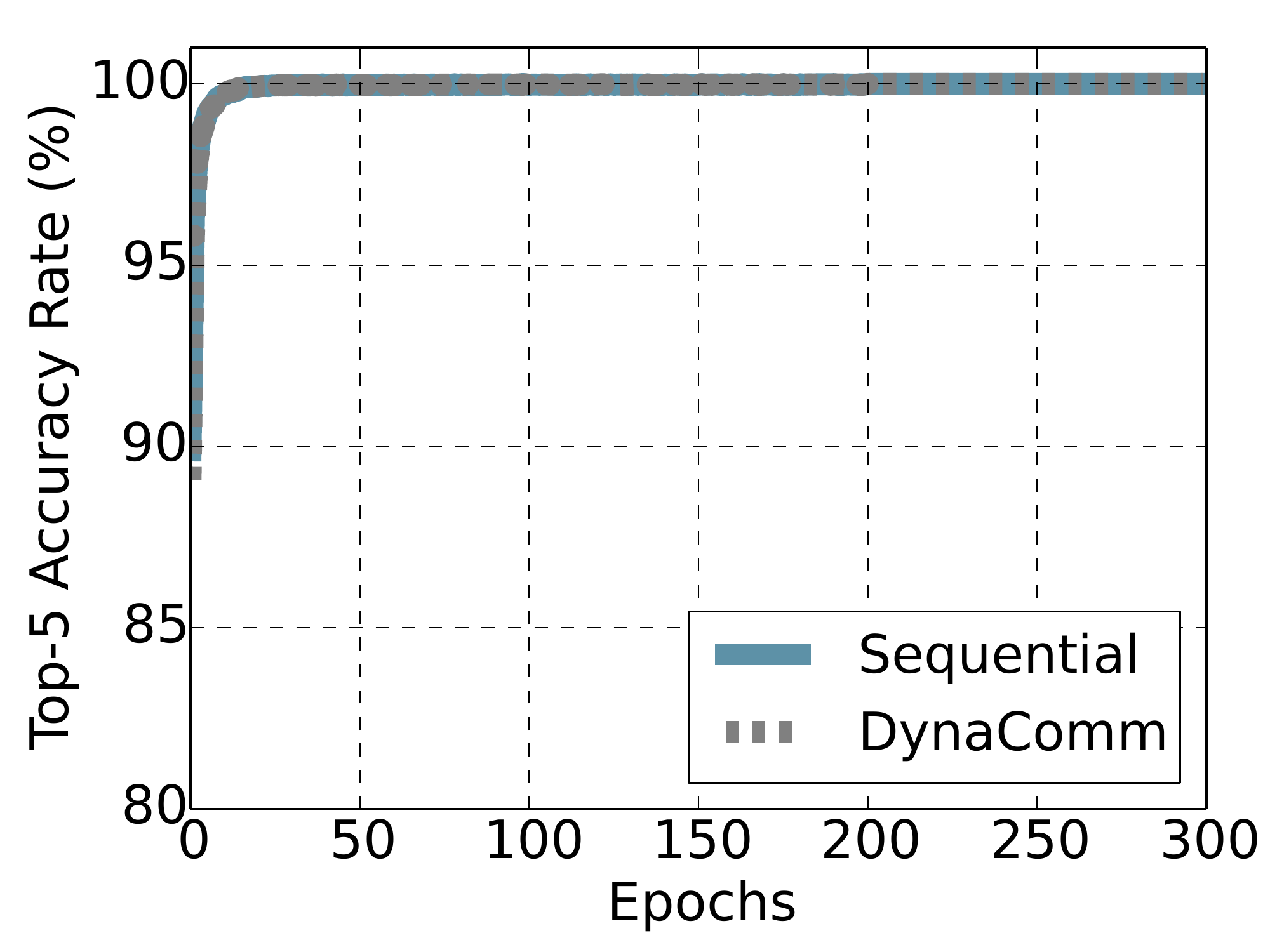}
%\caption{fig2}
\end{minipage}%
}%

\subfigure[Top-1 validation accuracy]{
\begin{minipage}[t]{0.475\linewidth}
\centering
\includegraphics[width=122px]{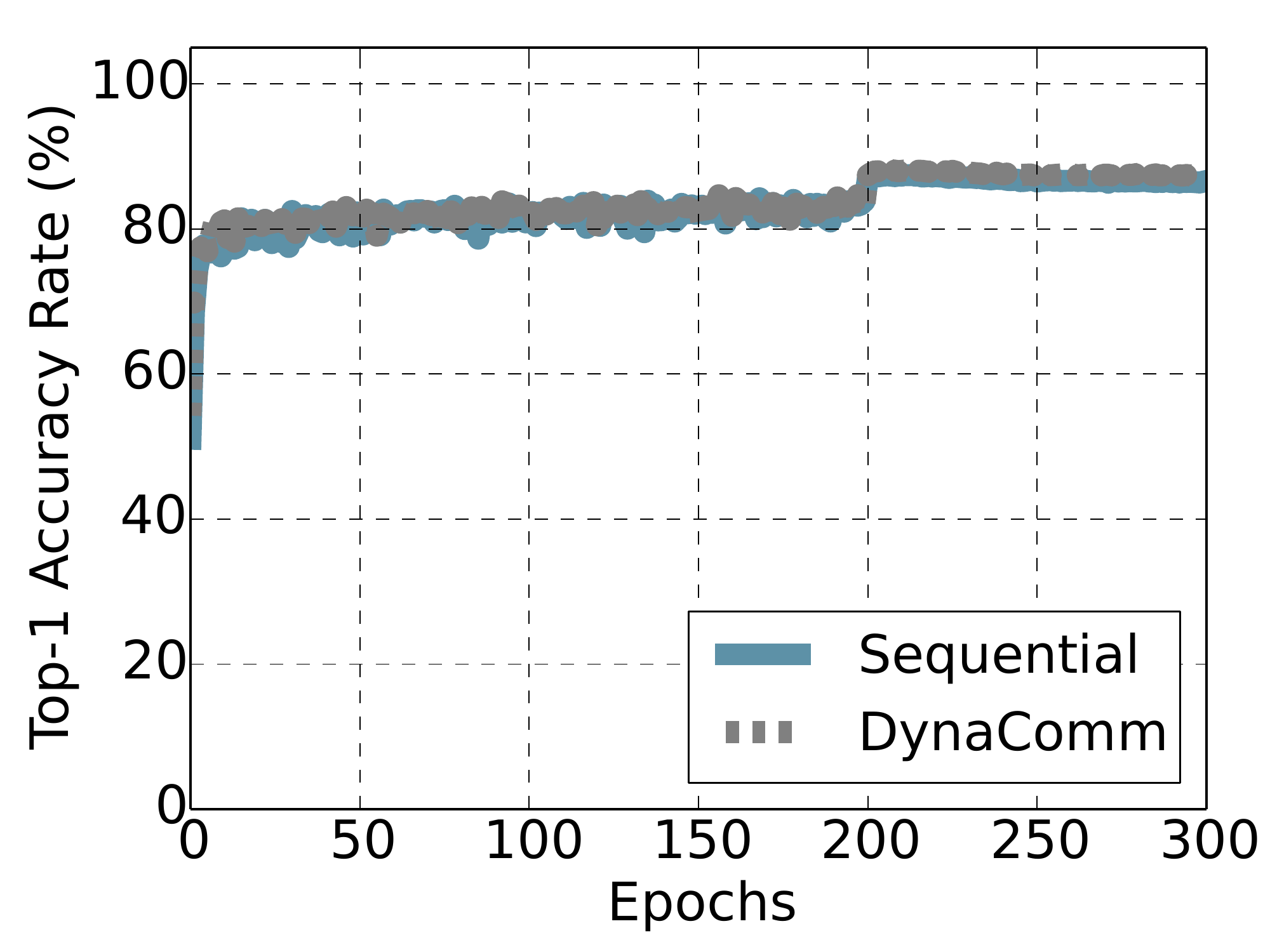}
%\caption{fig3}
\end{minipage}
}%
\subfigure[Top-5 validation accuracy]{
\begin{minipage}[t]{0.475\linewidth}
\centering
\includegraphics[width=122px]{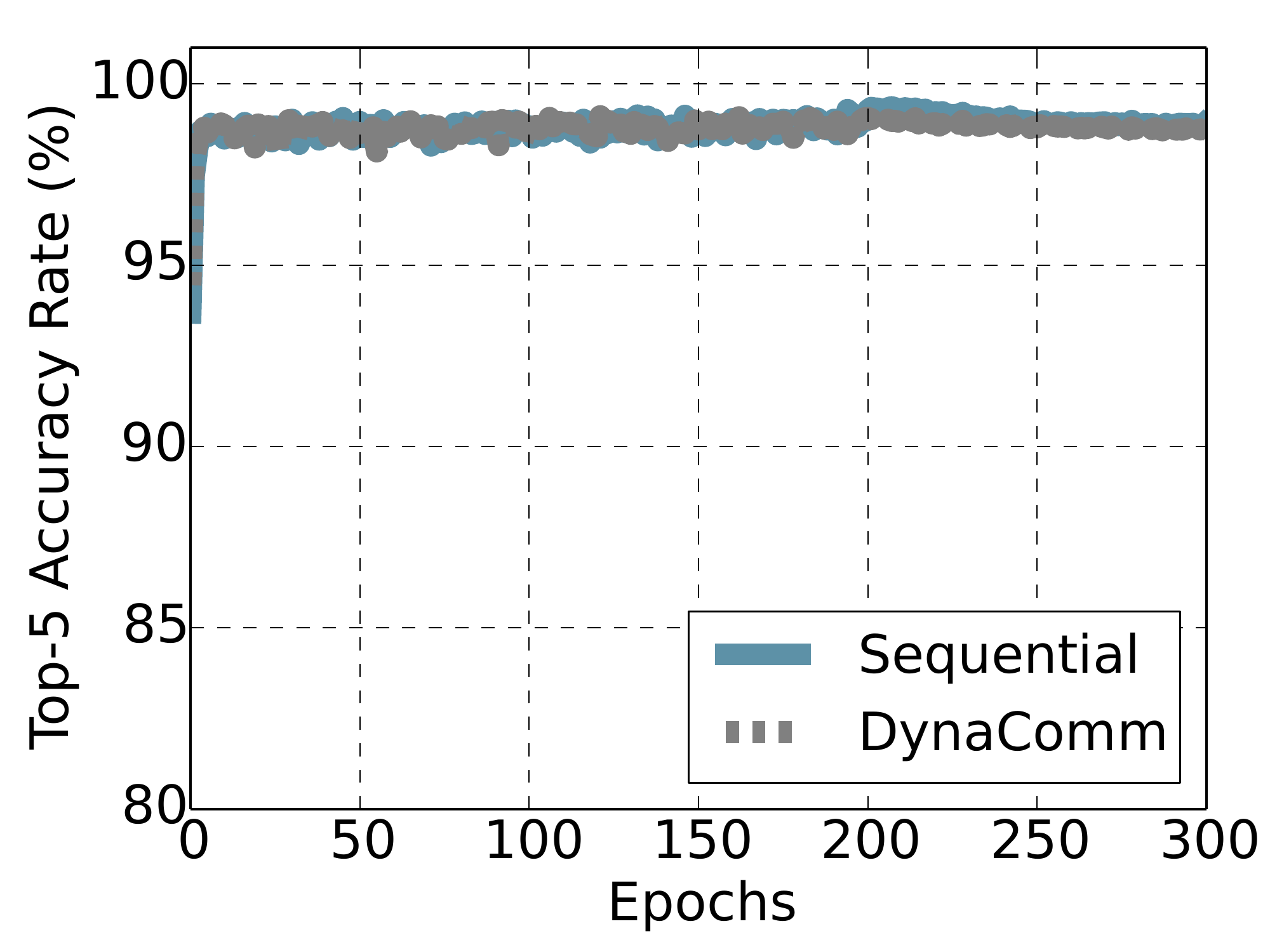}
%\caption{fig4}
\end{minipage}
}%

\centering
\caption{Top-1 and top-5 accuracy versus epoch with ResNet-152 on the CIFAR-10 dataset.}
\end{figure}

\begin{table*}[h]
\caption{Scheduling Overhead of DynaComm and iBatch}
\label{tab:scheduling_overhead}
\begin{center}
\begin{tabular}{ccccccc}
\hline
\textbf{Network} &\textbf{DynaComm/Fwd (ms)} &\textbf{iBatch/Fwd (ms)} &\textbf{$\Delta t + gt_i^1$ (ms)} &\textbf{DynaComm/Bwd (ms)} &\textbf{iBatch/Bwd (ms)} & \textbf{$\Delta t + pt_{i+1}^1$ (ms)}\\
\hline
VGG-19 & 0.084 $\pm$ 0.0027 & 0.30 $\pm$ 0.0099 & 14.06 $\pm$ 1.33 & 0.082 $\pm$ 0.0024 & 0.12 $\pm$ 0.0012 & 13.94 $\pm$ 1.40\\
GoogLeNet & 0.14 $\pm$ 0.0030 & 0.36 $\pm$ 0.0065 & 14.19 $\pm$ 1.52 & 0.15 $\pm$ 0.019 & 0.21 $\pm$ 0.0062 & 14.03 $\pm$ 1.56\\
Inception-v4 & 2.03 $\pm$ 0.0068 & 3.24 $\pm$ 0.019 & 13.93 $\pm$ 1.40 & 1.95 $\pm$ 0.012 & 1.85 $\pm$ 0.050 & 13.88 $\pm$ 1.36\\
ResNet-152 & 11.59 $\pm$ 0.080 & 12.65 $\pm$ 1.51 & 14.12 $\pm$ 1.72 & 11.56 $\pm$ 0.065 & 8.92 $\pm$ 1.08 & 14.17 $\pm$ 1.61\\
\hline
\end{tabular}
\end{center}
\end{table*}

\begin{figure}[t]
\centering
\includegraphics[scale=0.44]{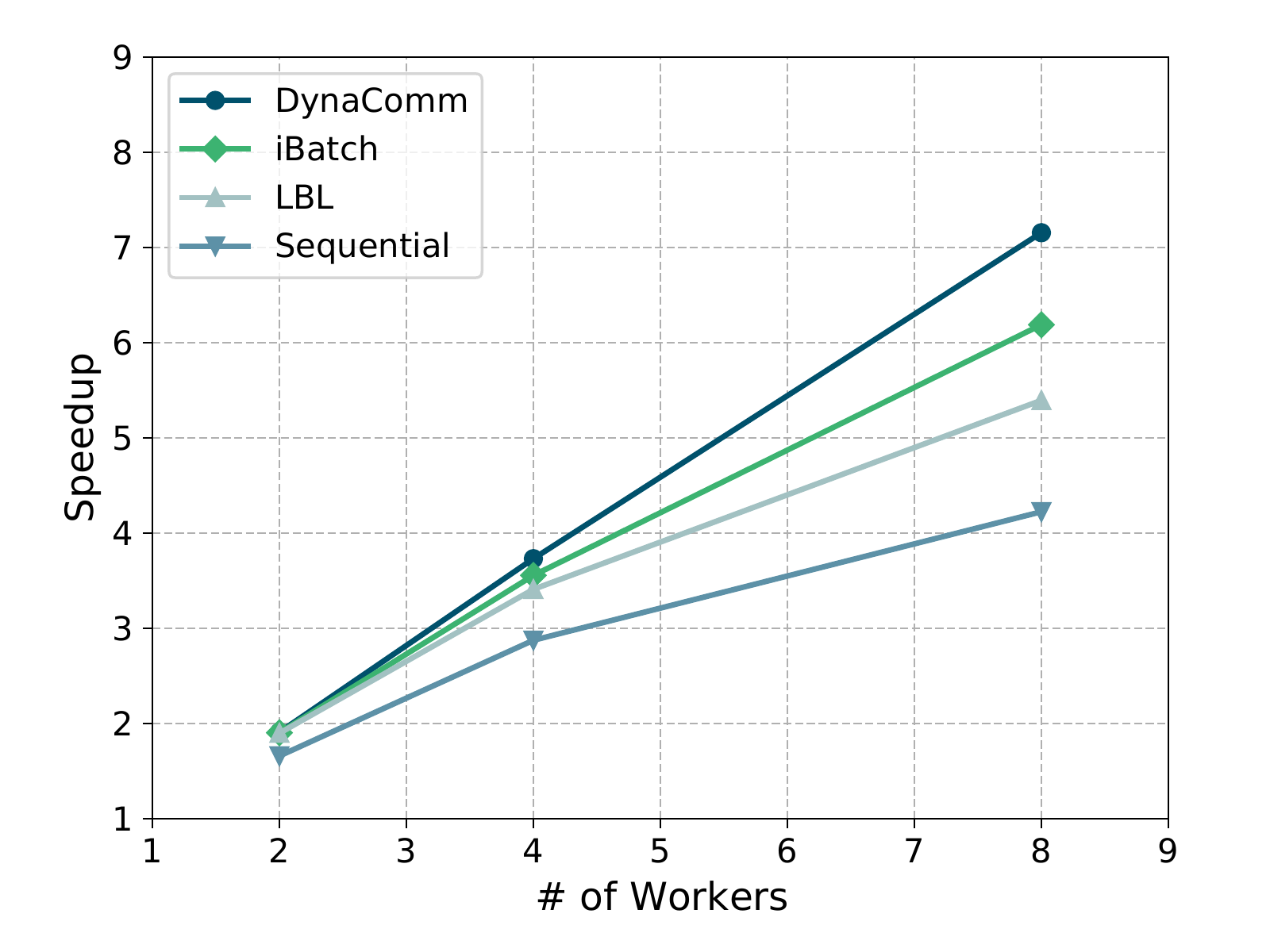}
\caption{Speedup versus the number of workers with ResNet-152 on the ILSVRC12 dataset.}
\label{fig:speedup_vs_workers}
\end{figure}

\textbf{Untouched Model Accuracy.} To verify that the layer-wise communication scheduling does not violate the inter-layer data dependency and the intra-layer procedure execution order, we trained ResNet-152 with and without DynaComm, respectively.

Fig. 10 illustrates the top-1 and top-5 accuracy of ResNet-152 on the CIFAR-10 dataset. To be mentioned, the top-1 accuracy examines the percentage of the model top-1 prediction output exactly matching the expected label, and the top-5 accuracy represents the percentage of any of the top-5 prediction outputs matching the expected label. Apart from this, the training accuracy means the model accuracy on the test dataset while the validation accuracy means the accuracy on the validation dataset. Compared to the original sequential implementation, our layer-wise communication scheduling method, DynaComm, affects neither the training accuracy, as is shown in Fig. 10 (a) and Fig. 10 (b), nor the validation accuracy, which is depicted in Fig. 10 (c) and Fig. 10 (d). Moreover, the training convergence property is also untouched according to these results. To be noticed, the layer-by-layer transmission strategy and iBatch are layer-wise communication scheduling strategies as well. To preserve the readability of Fig. 10, we only exhibit the results of DynaComm and the sequential execution scheme in the default PS (denoted as Sequential).

\textbf{Scalability.} In this subsection, we varied the number of workers to study the system scalability with ResNet-152 on the ILSVRC12 dataset. As exhibited in Fig. 11, when the cluster size is small, the communication traffic at the server-side is not heavy, and the communication cost is small compared to the computation cost. Therefore, all communication scheduling methods both achieve similar performance gains compared to the naive sequential implementation. However, When the number of workers increases, the network traffic also becomes heavier. Therefore, the performance gains brought by different communication scheduling methods start to diverge. From Fig. 11, we can observe that the system scalability with DynaComm is much better than that with other competing strategies when the number of workers grows. For instance, DynaComm achieves 7.2x speedup with 8-workers while iBatch achieves 6.2x speedup and the layer-by-layer transmission strategy achieves 5.4x speedup. Note that if the parameter servers finally get congested when the number of workers grows, the system maintainer should also scale out the server-side, or the saturated bandwidth and the communication costs will bottleneck the whole system.

\textbf{Minimizing Scheduling Overhead.} To study the overhead of the scheduling itself, we recorded the overall time costs while conducting different scheduling algorithms. We use \textit{DynaComm/Fwd} to represent the forward scheduling overhead and \textit{DynaComm/Bwd} to represent the backward scheduling overhead of DynaComm. Likewise, \textit{iBatch/Fwd} and \textit{iBach/Bwd} are the scheduling overheads of iBatch. Apart from this, \textit{$(\Delta t + gt_i^1)$} is the gradient transmission overhead of iteration $i$'s final layer while $(\Delta t + pt_{i+1}^1)$ is the first layer's parameter transmission cost of iteration $i+1$. Table I presents the mean value and the standard deviation of these data mentioned above. All results are averages of 5 runs. As is shown in Table I, the forward scheduling overheads of DynaComm and iBatch are less than the gradient transmission overhead \textit{$(\Delta t + gt_i^1)$} of iteration $i$. Therefore, DynaComm can launch the scheduling algorithm for the forward propagation in advance to minimize the scheduling overhead as mentioned in Section IV-C. Similarly, the backward scheduling algorithms can also be launched while the edge device is waiting for the first layer's parameter transmission of iteration $i+1$. Even if the overhead of the scheduling algorithms might be more expensive than the idle time window as the upcoming models get deeper and larger in the future, the scheduling overheads are reduced compared to the original and previous implementations. Also, the performance gains will far outweigh the introduced scheduling overheads.

\begin{table}[h]
\caption{Training Speed per worker with Profiling Switch on and off}
\label{tab:scheduling_overhead}
\begin{center}
\begin{tabular}{ccc}
\hline
\textbf{Network} &\textbf{On (samples/sec)} &\textbf{Off (samples/sec)}\\
\hline
VGG-19 & 4.46 $\pm$ 0.057 & 4.50 $\pm$ 0.046 \\
GoogLeNet & 29.62 $\pm$ 0.18 & 30.02 $\pm$ 0.27\\
Inception-v4 & 4.56 $\pm$ 0.038 & 4.62 $\pm$ 0.048\\
ResNet-152 & 4.48 $\pm$ 0.016 & 4.54 $\pm$ 0.018\\
\hline
\end{tabular}
\end{center}
\end{table}

\begin{figure}[t]
\centering

\subfigure[Forward scheduling overhead versus the number of layers.]{
\begin{minipage}[t]{\linewidth}
\centering
\includegraphics[width=210px]{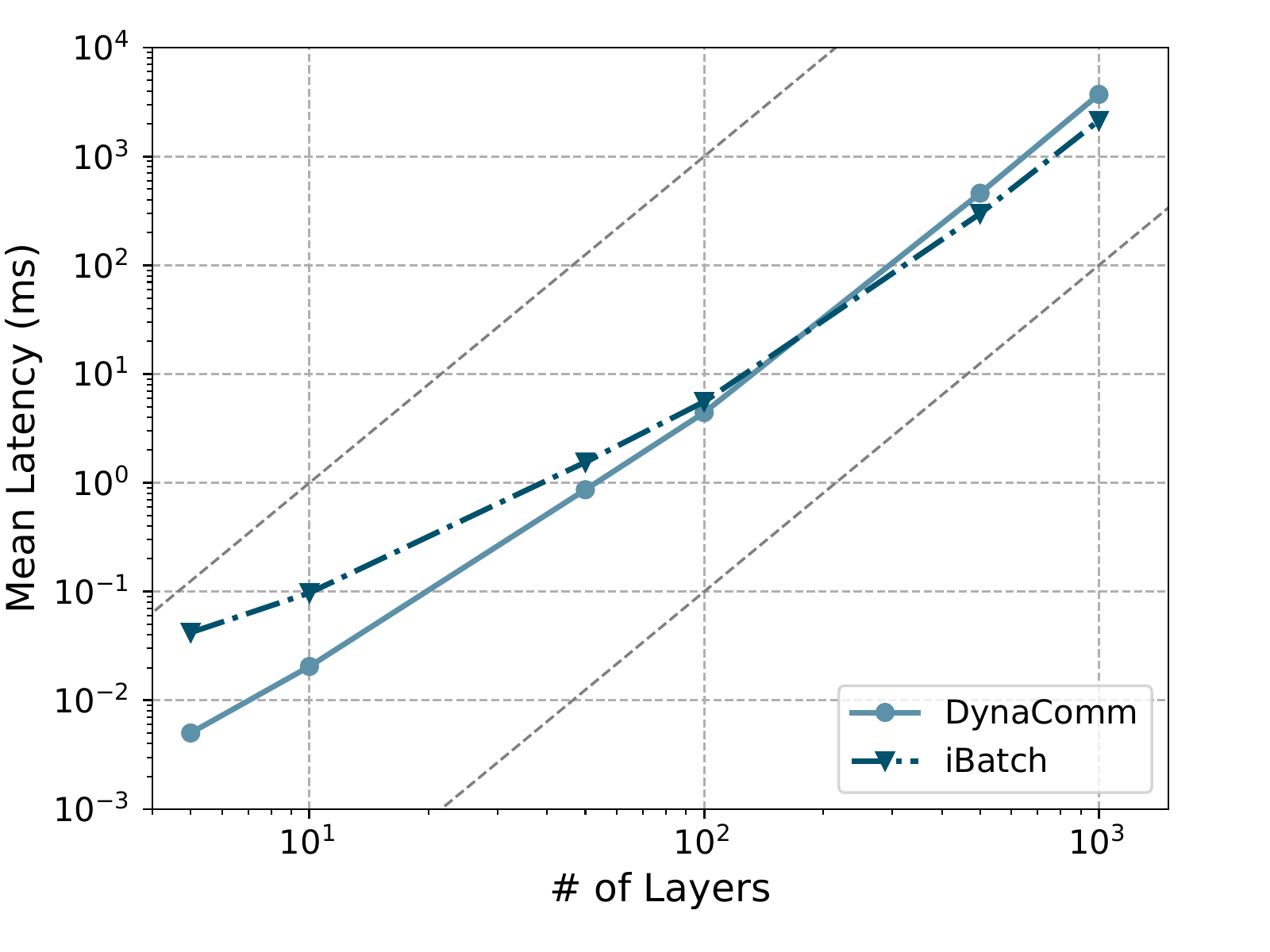}
%\caption{fig3}
\end{minipage}
}%

\subfigure[Backward scheduling overhead versus the number of layers.]{
\begin{minipage}[t]{\linewidth}
\centering
\includegraphics[width=210px]{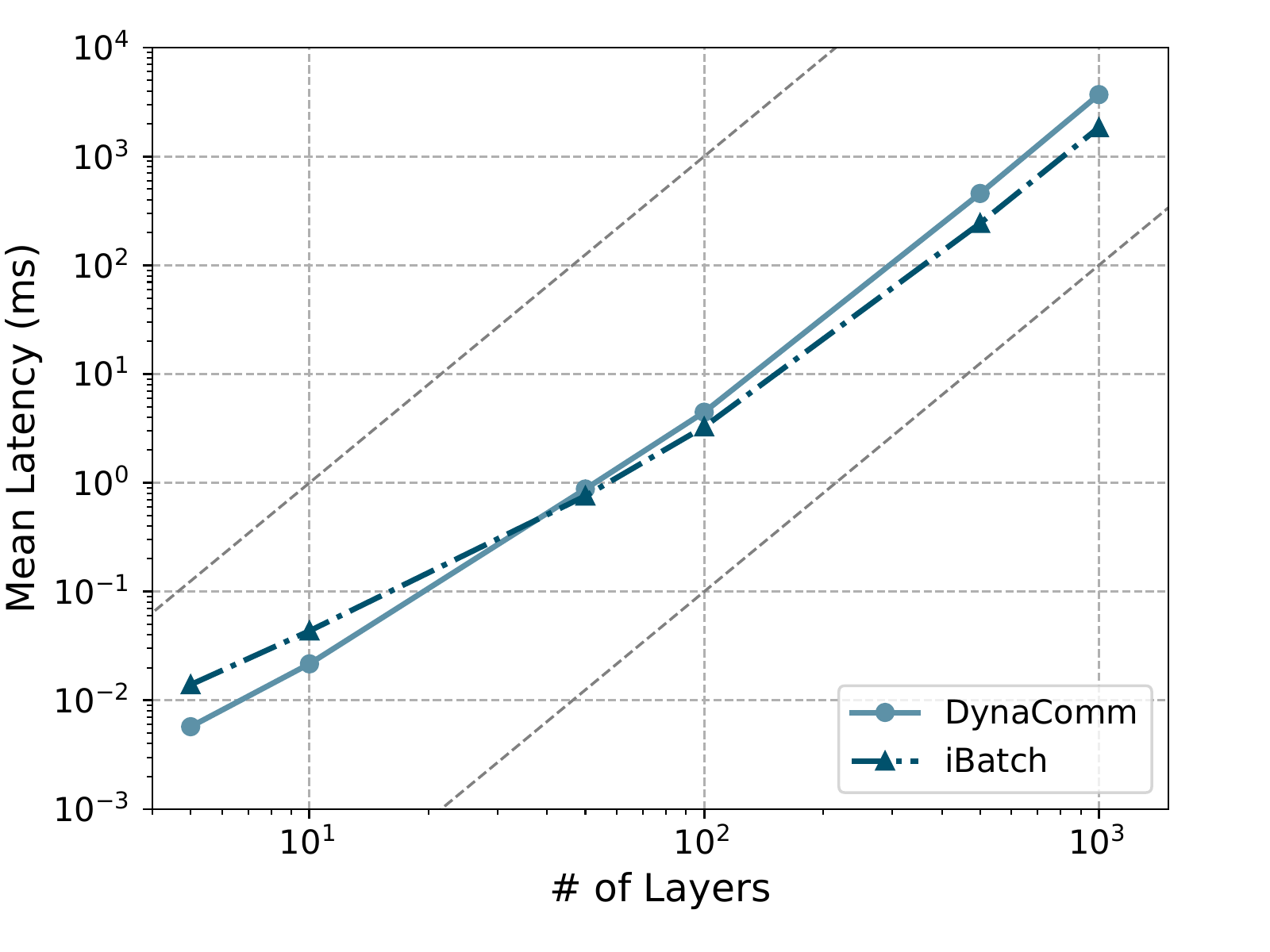}
%\caption{fig4}
\end{minipage}
}%

\centering
\caption{Scheduling overhead versus the number of network layers on generated profiling results.}
\end{figure}

Table II presents the mean value and the standard deviation of the local training speed with the profiling switch on and off. All results are averages of 5 runs. Although the real-time profiling processes also bring some extra computation overheads, these overheads are negligibly small that the highest local performance loss is only 1.33\% as shown in Table II.

Furthermore, to study the time complexity of these scheduling algorithms, we randomly generated a series of profiling results with different numbers of network layers. Then we recorded the overall time costs while conducting DynaComm's and iBatch's scheduling algorithms on these generated results. As is shown in Fig. 12, the time complexity of DynaComm is $\mathcal{O}(L^3)$ as analyzed in Section IV-B. In addition, we can also observe that DynaComm is more time-efficient than iBatch for the networks that are not deeper than 160 layers for the forward scheduling. As for the backward scheduling, the intersection is nearby 40, which is shown in Fig. 12 (b). When the targeted network goes deeper, iBatch takes less time because it settles for acceptable scheduling, not optimal scheduling, by conducting a greedy algorithm without the optimal substructure guarantee. Moreover, iBatch's algorithms have too many enumeration operations, which does not give it much of an advantage in terms of time complexity compared to DynaComm. With just a little more scheduling overhead, DynaComm manages to yield higher performance improvements when the targeted networks go deeper.

\section{Discussion}
In this section, we briefly discuss the applicability and limitations of DynaComm.

This layer-wise communication scheduling method we proposed applies only to layered models such as multi-layer perceptron and convolutional neural networks. It is based on the premise that each layer's parameters and gradients can be transmitted to the edge devices layer by layer without violating the computation dependency. Therefore, DynaComm and its competing methods are not applicable to some deep learning models such as recurrent neural networks.

Another limitation of all communication scheduling methods is that these techniques only exploit the potential performance gains by overlapping communications and computations. It means that there is an upper bound on the performance gains, which is very much related to the computation/communication ratio. As is mentioned in Section V, when the proportion of computation dominates, even though all communication overheads are hidden, the performance gains might still be trivial. Similarly, if the communication proportion is way larger than the computation proportion, the upper bound of the performance gains will be restricted to the computation proportion. All communication scheduling methods can be effective if and only if neither computation nor communication is a serious bottleneck to be addressed.

To achieve best practices in product environments, the system maintainer should (1) conduct local computation optimizations such as \cite{DBLP:conf/osdi/ZhengJSWYHWYZSG20,DBLP:conf/sosp/JiaPTWZA19} or reduce the workloads if computing capability is the bottleneck, and (2) upgrade the network interface card and other network hardware devices if network bandwidth and traffic are the major bottlenecks (or use gradient quantization techniques \cite{DBLP:conf/nips/WenXYWWCL17,DBLP:conf/nips/AlistarhG0TV17} if minor loss of accuracy is acceptable), then (3) choose an appropriate batch size through a few quick trials to balance the computation/communication ratio, and finally (4) conduct communication scheduling methods to minimize the iteration running time and improve system scalability.

\section{Conclusion}
In this paper, we present a communication scheduling problem, which is the underlying reason for the longer iteration execution time when performing deep learning at the network edge. To address the scheduling problem for both the parameters and the gradients communications for CNN training over edge networks, we present DynaComm, a novel DP-based communication scheduler, to conduct layer-wise communication scheduling during run-time. Through experiments, we verify that DynaComm manages to achieve optimal layer-wise scheduling compared to competing strategies, and it reduces the running time of each iteration by up to 41.92\% while the model accuracy remains untouched.

\section*{Acknowledgment}
We thank Tongliang Li, Yunren Bai, Airan Shao, and Ningxuan Feng for their insightful suggestions. This work is partially sponsored by National Key R\&D Program of China (No. 2019YFB2101700, 2018YFB0804402), National Science Foundation of China (U1736115), the Key Research and Development Project of Sichuan Province (No.21SYSX0082).

\ifCLASSOPTIONcaptionsoff
  \newpage
\fi



%\bibliographystyle{IEEEtran}

\bibliographystyle{IEEEtran}
\bibliography{dynacomm}


% % insert where needed to balance the two columns on the last page with
% %\newpage



%\vfill


\begin{IEEEbiography}[{\includegraphics[width=1in,height=1.25in,clip,keepaspectratio]{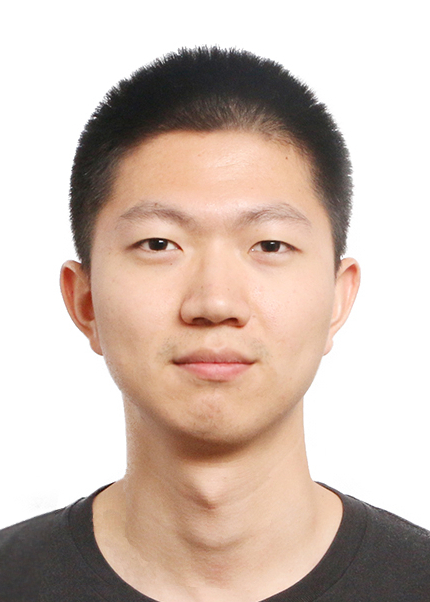}}]{Shangming Cai} received the B.S. degree in computer science from the College of Computer Science, Harbin Institute of Technology, China, in 2016. He is currently pursuing the Ph.D. degree in computer science with the Department of Computer Science and Technology, Tsinghua University, China. His main research interests include distributed machine learning, distributed storage, big data analytics, and edge computing.
\end{IEEEbiography}

\begin{IEEEbiography}[{\includegraphics[width=1in,height=1.25in,clip,keepaspectratio]{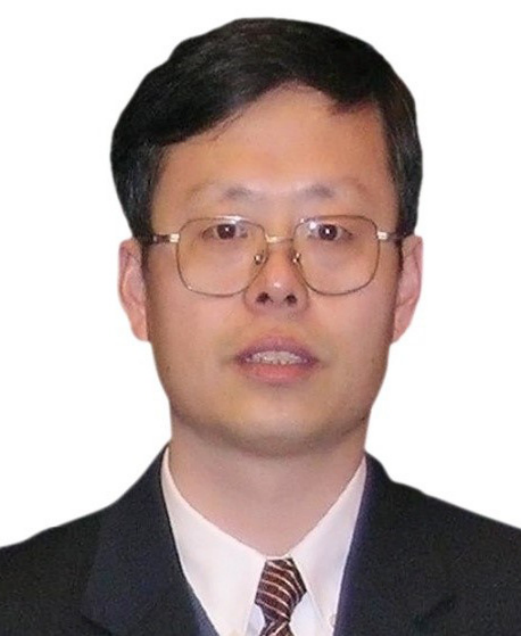}}]{Dongsheng Wang} (Member, IEEE) received the B.S., M.S., and Ph.D. degrees in computer science from Harbin Institute of Technology, Harbin, China, in 1989, 1992, and 1995, respectively. He is currently a Professor with the Department of Computer Science and Technology and Beijing National Research Center for Information Science and Technology, Tsinghua University, Beijing, China. Besides, he is also a Guest Professor with Cyberspace Security Research Center, Peng Cheng Laboratory, China. His research interests include computer architecture, high-performance computing, big data processing, and system security.
\end{IEEEbiography}

\begin{IEEEbiography}[{\includegraphics[width=1in,height=1.25in,clip,keepaspectratio]{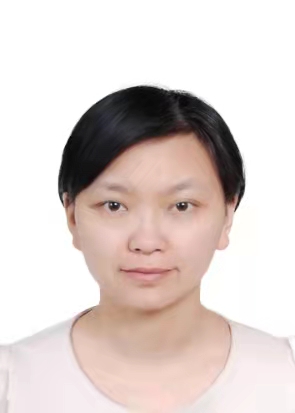}}]{Haixia Wang} (Member, IEEE) is an associate professor at the Beijing National Research Center for Information Science and Technology, Tsinghua University. She was born in 1977 and received her B.E. from Nankai University in 1998 and her Ph.D. degrees from Chinese Academy of Sciences in 2004. Her major research interests include microprocessor architecture, distributed systems, and fault tolerance.
\end{IEEEbiography}

\begin{IEEEbiography}[{\includegraphics[width=1in,height=1.25in,clip,keepaspectratio]{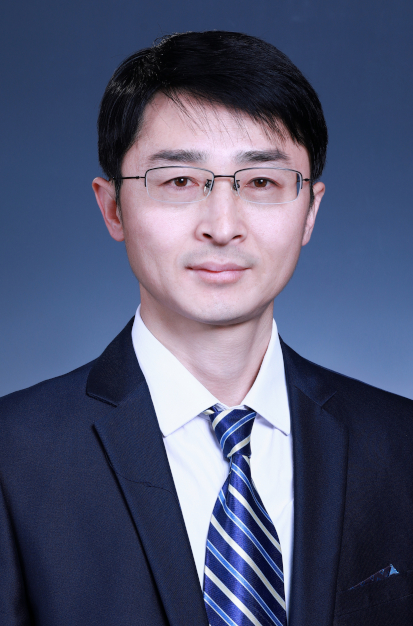}}]{Yongqiang Lyu} (Member, IEEE) received the B.S. degree in computer science from Xidian University, Xi’an, China, in 2001, and the M.S. and the Ph.D. degrees in computer science from Tsinghua University, Beijing, China, in 2003 and 2006 respectively. He is currently an Associate Professor with the National Research Center for Information Science and Technology, Tsinghua University. His research interests focus on processor hardware security, computer system security, networking, and the IoTs.
\end{IEEEbiography}

\begin{IEEEbiography}[{\includegraphics[width=1in,height=1.25in,clip,keepaspectratio]{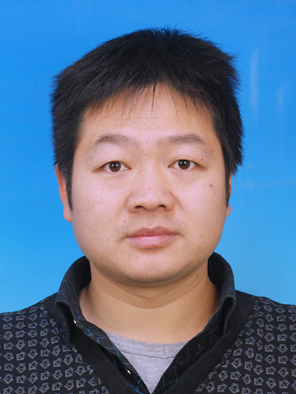}}]{Guangquan Xu} (Member, IEEE) is a Ph.D. and full professor at the Tianjin Key Laboratory of Advanced Networking (TANK), College of Intelligence and Computing, Tianjin University, China. He received his Ph.D. degree from Tianjin University in March 2008. His research interests include cyber security and trust management. He is the director of Network Security Joint Lab and the Network Attack Defense Joint Lab. He has published 100+ papers in reputable international journals and conferences, including IEEE Transactions on Cybernetics, IEEE Internet of Things Journal, ACM Transactions on Internet Technology, ACM Transactions on Intelligent Systems and Technology, IEEE Transactions on Industrial Informatics, Information Sciences, IEEE Wireless Communications, IEEE Network, Computers Security, and so on. He served as a TPC member for IEEE UIC 2018, SPNCE2019, IEEE UIC2015, IEEE ICECCS 2014, and reviewers for journals such as IEEE Access, ACM TIST, JPDC, IEEE TITS, soft computing, FGCS, and Computational Intelligence, and so on. He is a Fellow of IET, IEEE member, senior member of China Computer Society.
E-mail: losin@tju.edu.cn https://orcid.org/0000-0001-8701-3944
\end{IEEEbiography}

\begin{IEEEbiography}[{\includegraphics[width=1in,height=1.25in,clip,keepaspectratio]{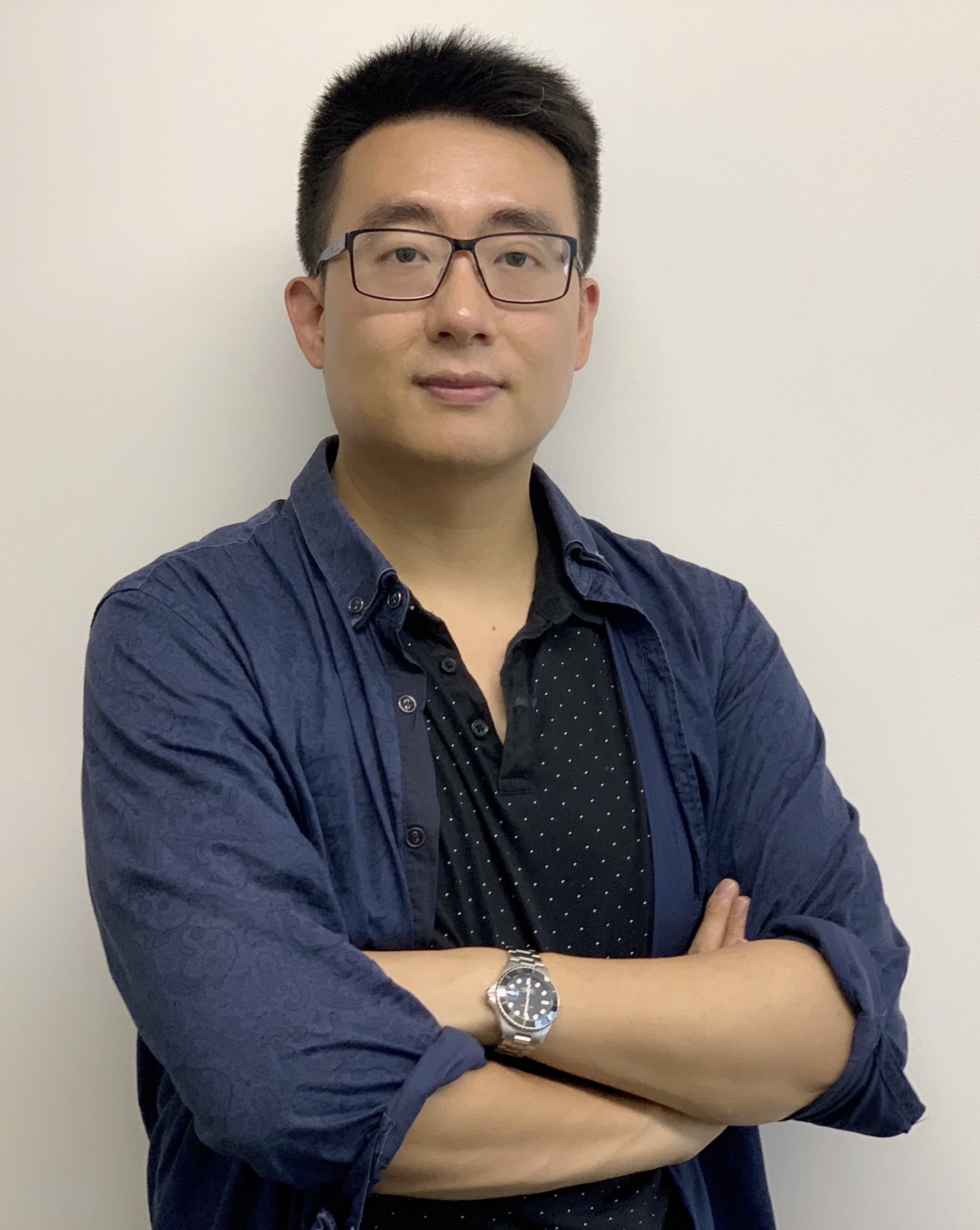}}]{Xi Zheng} (Member, IEEE) got Ph.D. in Software Engineering from UT Austin. He specialized in Machine Learning Testing, Distributed learning and Embedded Intelligence, IoT Security, and Reliability Analysis. Now Director of Intelligent Systems Research Group and Associate Professor/Senior Lecturer in Software Engineering at Macquarie University. Published more than 80 high-quality publications in top journals and conferences. PC for PerCom and TrustCom. Awarded the best paper in Australian distributed computing and doctoral conference in 2017. Awarded Deakin Research outstanding award in 2016 and Macquarie Early Career Research Highly Commended in 2020. Awarded Multiple ARC LP and DP projects. Active reviewer for top journals and conferences.
\end{IEEEbiography}

\begin{IEEEbiography}[{\includegraphics[width=1in,height=1.25in,clip,keepaspectratio]{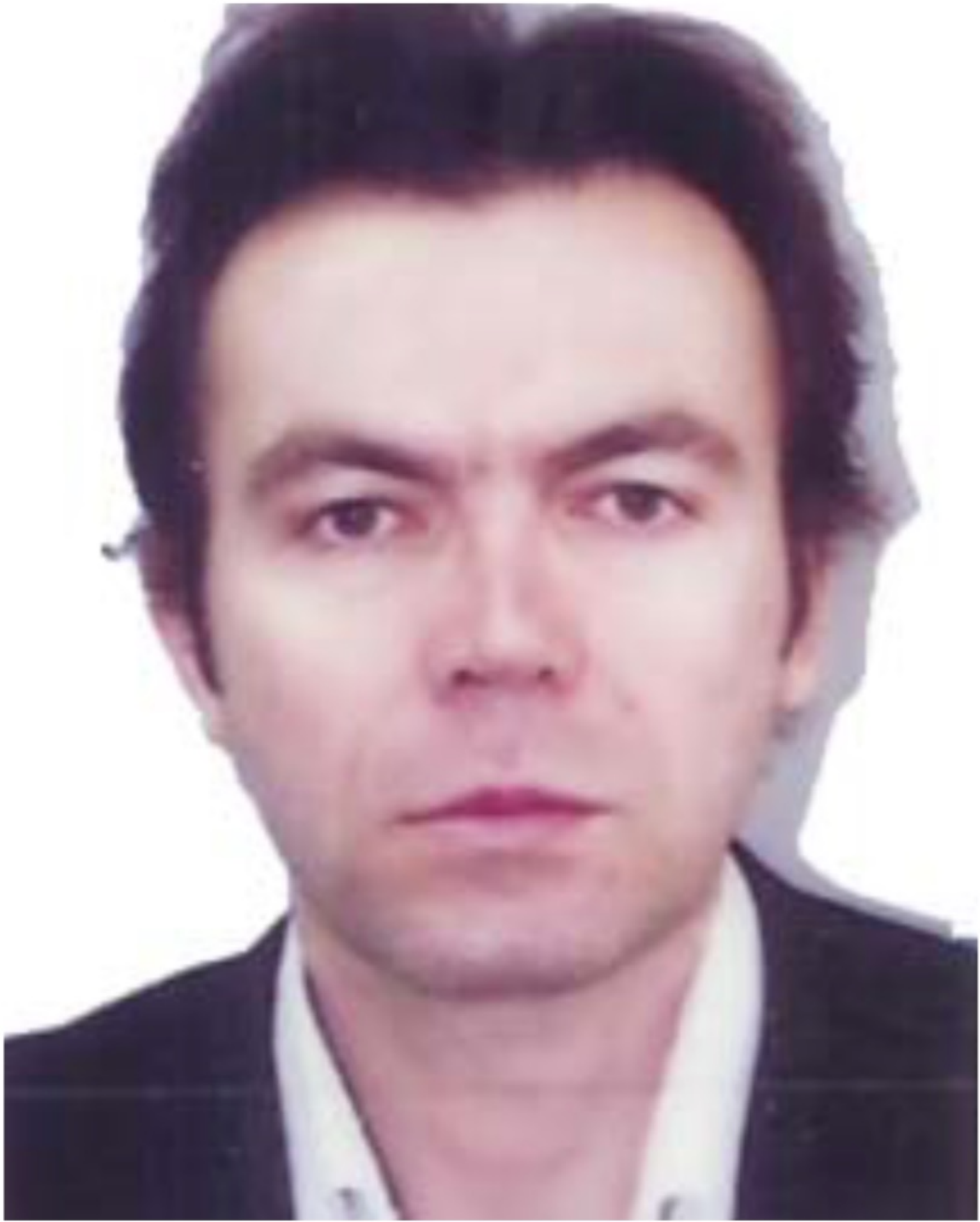}}]{Athanasios V. Vasilakos} (Senior Member, IEEE) received the Ph.D. degree in computer engineering from the University of Patras, Patras, Greece, in 1988. He is currently with the College of Mathematics and Computer Science, Fuzhou University, China, with the School of Electrical and Data Engineering, University of Technology Sydney, Australia, and with the Department of Computer Science, Electrical and Space Engineering, Lulea University of Technology, Sweden. He has authored or co-authored 600 papers in peer-reviewed journals and conferences and is a WoS Highly Cited Researcher with more than 52000 citations and H-Index=120. His main research interests include cybersecurity, networking, the IoTs, and big data analytics. Prof. Vasilakos is an Editor for many technical journals, such as the IEEE Transactions on Network and Service Management, the IEEE Transactions on Cloud Computing, the IEEE Transactions on Information Forensics and Security, the IEEE Transactions on Cybernetics, the IEEE Transactions on NanoBioscience, and the ACM Transactions on Autonomous and Adaptive Systems. He was the General Chair of the European Alliances for Innovation. (E-mail:th.vasilakos@gmail.com)
\end{IEEEbiography}

\end{document}